\documentclass[sigconf,screen]{acmart}

\usepackage[linesnumbered,ruled]{algorithm2e}
\makeatletter
\renewcommand{\@algocf@capt@plain}{above}
\makeatother
\usepackage{algorithmicx}
\usepackage{algpseudocode}    
\usepackage{flushend}
\usepackage{amsmath}
\usepackage{bbding}
\usepackage{booktabs}
\usepackage{calc}
\usepackage{circuitikz}
\usepackage{color}            
\usepackage{etex}
\usepackage{graphicx}
\usepackage{listings}
\usepackage{mdwlist}          
\usepackage{microtype}        
\usepackage{multirow}
\usepackage{paralist} 
\usepackage{relsize}
\usepackage{setspace}
\usepackage{subfigure}
\usepackage[T1]{fontenc}      
\usepackage{tikz}
\usepackage[utf8]{inputenc}   
\usepackage[USenglish]{babel} 
\usetikzlibrary{matrix,arrows,positioning,shapes,backgrounds,fit}

\newcommand\ignore[1]{}

\newcommand{\textcode}[1]{\texttt{#1}}

\newcommand{\mem}{\mathcal{M}}
\newcommand{\pcon}{\mathit{pcon}}

\newcommand{\outline}[1]{}

\newcommand{\SymSC}{\textsc{SymSC}}

\let\oldnl\nl
\newcommand{\nonl}{\renewcommand{\nl}{\let\nl\oldnl}}

\definecolor{darkblue}{rgb}{0, 0.125, 0.576}
\definecolor{dkgreen}{rgb}{0,0.6,0}
\definecolor{gray}{rgb}{0.5,0.5,0.5}
\definecolor{mauve}{rgb}{0.58,0,0.82}

\newcommand{\submissioncomment}[1]{  }

\definecolor{salmon}{RGB}{255,191,191}
\tikzset{
  woval/.style={minglingh
    draw
    , line width=1pt
    , anchor=center
    , text centered
    , rounded corners
  },
}
\tikzset{
  hoval/.style={
    draw
    , line width=1pt
    , fill=salmon
    , anchor=center
    , text centered
    , rounded corners
  },
}



\setcopyright{none}
\acmDOI{}
\acmISBN{}
\acmYear{}
\copyrightyear{}
\acmPrice{}
\acmConference[]{}{}{}

\begin{document}

\setlength{\abovedisplayskip}{3pt}
\setlength{\belowdisplayskip}{3pt}

\title[]{Adversarial Symbolic Execution for Detecting Concurrency-Related Cache Timing Leaks}

\author{Shengjian Guo}
\affiliation{%
  \institution{Virginia Tech}
  \city{Blacksburg}
  \state{VA}
  \postcode{24061}
  \country{USA}
}

\author{Meng Wu}
\affiliation{%
  \institution{Virginia Tech}
  \city{Blacksburg}
  \state{VA}
  \postcode{24061}
  \country{USA}
}

\author{Chao Wang}
\affiliation{%
  \institution{University of Southern California}
  \city{Los Angeles}
  \state{CA}
  \postcode{90089}  
  \country{USA}
}

\begin{abstract}
The timing characteristics of cache, a high-speed storage between the
fast CPU and the slow memory, may reveal sensitive information of a
program, thus allowing an adversary to conduct side-channel attacks.
Existing methods for detecting timing leaks either ignore cache all
together or focus only on passive leaks generated by the program
itself, without considering leaks that are made possible by
concurrently running some other threads.  In this work, we show that
\emph{timing-leak-freedom} is not a compositional property: a program
that is not leaky when running alone may become leaky when interleaved
with other threads.
Thus, we develop a new method, named \emph{adversarial symbolic
  execution}, to detect such leaks.  It systematically explores both
the feasible program paths and their interleavings while modeling the
cache, and leverages an SMT solver to decide if there are timing
leaks.
We have implemented our method in LLVM and evaluated it on a set of
real-world ciphers with 14,455 lines of C code in total.  Our
experiments demonstrate both the efficiency of our method and its
effectiveness in detecting side-channel leaks.
\end{abstract}



\maketitle

\section{Introduction}
\label{sec:intro}

Side-channel attacks are security attacks where an adversary exploits
the dependency between sensitive data and non-functional properties of
a program such as the execution time~\cite{Kocher96,DhemKLMQW98},
power consumption~\cite{KocherJJ99,MangardOP07}, heat,
sound~\cite{GenkinST14}, and electromagnetic
radiation~\cite{GandolfiMO01,Quisquater2001}.
%
For timing side channels, in particular, there are two main sources of
leaks: variances in the number of executed instructions and variances
in the cache behavior.
Instruction-induced leaks are caused by differences in the number and
type of instructions executed along different paths: unless the
differences are independent of the sensitive data, they may be
exploited by an adversary.
Cache-induced leaks are caused by differences in the number of cache
hits and misses along different paths.

Existing methods for detecting timing leaks or proving their absence
often ignore the cache all together while focusing on
instruction-induced leaks.  For example, Chen et al.~\cite{ChenFD17}
used Cartesian Hoare Logic~\cite{SousaD16} to prove the timing leak of
a program is within a bound; Antonopoulos et
al.~\cite{AntonopoulosGHK17} used a similar technique that partitions
the set of program paths in a way that, if individual partitions are
proved to be timing attack resilient, the entire program is also
timing attack resilient.  Unfortunately, these methods ignore the
cache-timing characteristics.
Even for techniques that consider the
cache~\cite{KopfMO12,DoychevFKMR13,ChuJM16,BasuC17,TouzeauMMR17,Chattopadhyay17},
their focus has been on leaks manifested by the program itself when
running alone, without considering the cases when it is executed
concurrently with some other (benign or adversarial) threads.

In this work, we show \emph{side-channel leak-freedom}, as a security
property, is not compositional.  That is, a leak-free program when
running alone may still be leaky when it is interleaved with other
threads, provided that they share the memory subsystem.  This is the
case even if all paths in the program have the same number and type of
instructions and thus do not have \emph{instruction-induced} timing
leaks at all.  Unfortunately, no existing method or tool is capable of
detecting such timing leaks.

We propose a new method, named \emph{adversarial symbolic execution}, to
detect such concurrency-related timing leaks.
Specifically, given a program where one thread conducts a
security-critical computation, e.g., by calling functions in a
cryptographic library, and another thread is (either accidentally or
intentionally) adversarial, our method systematically explores both
paths in these threads and their interleavings.  The exploration is
symbolic in that it covers feasible paths under all input values.
During the symbolic execution, we aim to analyze the cache behavior
related to sensitive data to detect timing leaks caused by the
interleaving.

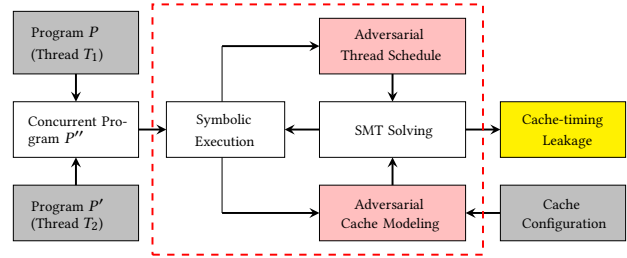
\begin{figure}
\vspace{2ex}
\centering
\scalebox{0.925}{
\definecolor{salmon}{RGB}{255,191,191}
\begin{tikzpicture}[fill=blue!20,font=\scriptsize] 
  \tikzstyle{arrow}=[thick,->,>=stealth,black]
  \tikzstyle{txt}=[above=5pt,right,text width=1.6cm]
  \tikzstyle{long_txt}=[above=5pt,right,text width=2cm]
  \tikzstyle{short_txt}=[above=5pt,right,text width=1.4cm]
  \draw [fill=lightgray](0, 0) rectangle (1.8, 0.8);
  \node [above=5pt,right,text width=1.6cm] at (0.15, 0.2) {Program $P'$ (Thread $T_2$)};
  \draw (0.0, 1.2) rectangle (1.8, 2.0);
  \node [above=5pt,right,text width=1.6cm] at (0.1 , 1.4) {Concurrent Program $P''$};
  \draw [fill=lightgray](0.0, 2.4) rectangle (1.8, 3.3);
  \node [txt] at (0.15,2.8){Program $P$};
  \node [txt] at (0.15,2.5){(Thread $T_1$)};
  \draw [arrow](0.9,0.8)--(0.9,1.2);
  \draw [arrow](0.9,2.4)--(0.9,2.0);
  \draw [arrow](1.8,1.6)--(2.2,1.6);
  \draw (2.2, 1.2) rectangle (3.9, 2.0);
  \node [txt] at (2.52,1.55){Symbolic};
  \node [txt] at (2.5,1.25){Execution};

  \draw (3.0,1.2)--(3.0,0.4);

  \draw [arrow](3.0,0.4)--(4.4,0.4);
  \draw [arrow](4.4,1.6)--(3.9,1.6);
  \draw [fill=salmon] (4.4, 0) rectangle (6.5, 0.8);
  \node [txt] at (4.8,0.35){Adversarial};
  \node [long_txt] at (4.6,0.05){Cache Modeling};

  \draw (4.4, 1.2) rectangle (6.5, 2.0);
  \node [long_txt] at (4.8,1.4){SMT Solving};
  \draw [arrow](5.45,0.8)--(5.45,1.2);

  \draw [arrow](7.0,0.4)--(6.5,0.4);
  \draw [arrow](6.5,1.6)--(7.0,1.6);
  \draw [fill=lightgray](7.0, 0) rectangle (8.8, 0.8);
  \node [short_txt] at (7.5,0.35){Cache};
  \node [short_txt] at (7.2,0.05){Configuration};
  \draw [fill=yellow](7.0, 1.2) rectangle (8.8, 2.0);
  \node [txt] at (7.2,1.55){Cache-timing};
  \node [txt] at (7.5,1.25){Leakage};
  \draw [fill=salmon] (4.4, 2.4) rectangle (6.5, 3.2);
  \node [txt] at (4.8, 2.8) {Adversarial};
  \node [long_txt] at (4.6, 2.5) {Thread Schedule};

  \draw  (3.0, 2.0)--(3.0, 2.8);
  \draw [arrow] (3.0, 2.8)--(4.4, 2.8);
  \draw [arrow] (5.45, 2.4)--(5.45, 2.0);
  \draw [dashed,red,thick](2.0, -0.2) rectangle (6.75, 3.4);
\end{tikzpicture}

}
\caption{Flow of our cache timing leak detector \SymSC{}.}
\label{fig:overall_flow}
\vspace{-1ex}
\end{figure}

Figure~\ref{fig:overall_flow} shows the flow of our leak detector
named \SymSC{}, which takes the victim thread $P$, a potentially
adversarial thread $P'$, and the cache configuration as input.
If $P'$ is not given, \SymSC{} creates it automatically.
While symbolically executing the program, \SymSC{} explores all thread
paths and searches for an adversarial interleaving of these paths
that exposes divergent cache behaviors in $P$.
There are two main technical challenges.
The first one is associated with systematic exploration of the
interleaved executions of a concurrent program so as not to miss any
adversarial interleaving.
The second one is associated with modeling the cache accurately while
reducing the computational cost.

To address the first challenge, we developed a new algorithm for
adversarially exploring the interleaved executions while mitigating
the \emph{path and interleaving explosions}.  Specifically, cache
timing behavior constraints, which are constructed \emph{on the fly}
during symbolic execution, are leveraged to prune interleavings
redundant for detecting leaks and thus speed up the exploration.

To address the second challenge, we developed a technique for modeling
the cache behavior of a program based on the cache's type and
configuration, as well as optimizations of the subsequent constraint
solving to reduce overhead.
For each concurrent execution (an interleaving of the threads) denoted
$\pi=(in,sch)$, where $in$ is the sensitive data input and $sch$ is
the interleaving schedule, we construct a logical constraint
$\tau_t(in,sch)$ for every potentially adversarial memory access $t$,
to indicate when it leads to a cache hit.  Then, we seek two distinct
values of the data input, $in$ and $in'$, for which the cache behaves
differently: $\tau_t(in,sch) \ne \tau_t(in',sch)$, meaning one of them
is a hit but the other is a miss, and they are due to differences in
the sensitive data input.

We have implemented our method in a software tool based on LLVM and
the KLEE symbolic virtual machine~\cite{CadarDE08}, and evaluated it
on twenty benchmark programs.  These security-critical programs are
ciphers taken from cryptographic libraries in the public domain; they
have 14,455 lines of C code in total.  Since these programs are
crafted by domain experts, they do not have obvious timing leaks when
running alone, such as unbalanced branching statements or variances in
lookup-table accesses.  However, our experiments of applying \SymSC{}
show that they may still have timing leaks when being executed
concurrently with other threads.

To summarize, we make the following contributions:
\begin{itemize}
\item 
We propose an \emph{adversarial symbolic execution} method capable of
detecting cache timing leaks in a security-critical program when it
runs concurrently with other threads.
\item 
We implement and evaluate our method on real-world cipher programs to
demonstrate its effectiveness in detecting concurrency-related 
timing leaks.
\end{itemize}

In the remainder of this paper, we first motivate our work using
several examples in Section~\ref{sec:motivation} and then provide the
technical background in Section~\ref{sec:prelim}.  We present our
detailed algorithms in Sections~\ref{sec:adversarialSE} and
\ref{sec:adversarialCM}, which are followed by domain-specific
optimizations in Section~\ref{sec:optimization} to reduce the
computational overhead.  We present our experimental results in
Section~\ref{sec:evaluation} and review the related work in
Section~\ref{sec:related}.  Finally, we give our conclusions in
Section~\ref{sec:conclusion}.

\section{Motivation}
\label{sec:motivation}

In this section, we use examples to explain the difference between
self-leaking and concurrency-induced leaking.

\begin{figure}
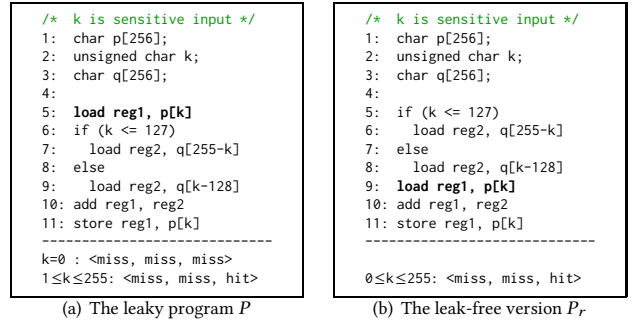

\vspace{1ex}
\centering
\subfigure[The leaky program $P$]{
  \label{fig:fig2-a}
  \framebox[.46\linewidth]{     
    \begin{minipage}{0.45\linewidth}
      {\scriptsize \tt
	\begin{tabbing}
	  xxx \= xxx \= \kill
	  \textcolor{dkgreen}{\texttt{/*}}\>\textcolor{dkgreen}{\texttt{k is sensitive input\ */}}    \\
	  1: \> char p[256];  	 	\\              
	  2: \> unsigned char k;  	\\              
	  3: \> char q[256];         	\\  
  	  4: \>          	\\  
	  5: \>  \textbf{load reg1, p[k]} 		\\
	  6: \>  if (k <= 127) 		\\
	  7: \>  	~ load reg2, q[255-k] \\
    	  8: \>  else \\
    	  9: \>  	~ load reg2, q[k-128] \\
    	  10: \> add reg1, reg2 \\
    	  11: \> store reg1, p[k] \\
    	  -----------------------------\\
    	  k=0 : <miss, miss, miss> \\
    	  1$\leq$k$\leq$255: <miss, miss, hit> 
	\end{tabbing}
      }
    \end{minipage}
  }
}
\hspace{.01\linewidth}
\subfigure[The leak-free version $P_r$]{
  \label{fig:fig2-b}
  \framebox[.46\linewidth]{     
    \begin{minipage}{0.45\linewidth}
      {\scriptsize \tt
	\begin{tabbing}
	  xxx \= xxx \= \kill
	  \textcolor{dkgreen}{\texttt{/*}}\>\textcolor{dkgreen}{\texttt{k is sensitive input\ */}}    \\
	  1: \> char p[256];  	 	\\              
	  2: \> unsigned char k;  	\\              
	  3: \> char q[256];         	\\  
  	  4: \>          	\\  
	  5: \>  if (k <= 127) 		\\
	  6: \>  	~ load reg2, q[255-k] \\
    	  7: \>  else \\
    	  8: \>  	~ load reg2, q[k-128] \\
    	  9: \>  \textbf{load reg1, p[k]} 		\\
    	  10: \> add reg1, reg2 \\
    	  11: \> store reg1, p[k]\\
    	  -----------------------------\\
    	  \\
    	  0$\leq$k$\leq$255: <miss, miss, hit> 
	\end{tabbing}
      }
    \end{minipage}
  }
}

\vspace{-2ex}
\caption{A program with cache-timing  leak (cf.~\cite{ChattopadhyayBRZ17}).}
\label{fig:motiv-seq}
\vspace{-2ex}
\end{figure}

\subsection{A Self-leaking Program and the Repair}
\label{sec:leakage-example}

Figure~\ref{fig:motiv-seq}(a) shows a program whose execution time is
dependent of the sensitive variable \texttt{k}.  It is a revised
version of the running example used in~\cite{ChattopadhyayBRZ17}, for
which the authors proposed the leak-free version shown in
Figure~\ref{fig:motiv-seq}(b).  The two programs have the same set of
instructions but differ in where the highlighted \texttt{load}
instruction is located: line 5 in $P$ and line 9 in $P_r$.

\begin{figure}
\vspace{1ex}
\centering
\scalebox{.9}{
\begin{tikzpicture}[fill=blue!20,font=\footnotesize] 
  \draw (0, 0) rectangle (2.2, 0.4);
  \node[above=5pt, right] at (0.6, 0) {q[254]};
  \draw[dotted] (0, 0.4) rectangle (2.2, 0.8);
  \node[above=5pt, right] at (0.7, 0.4) {......} ;	
  \draw (0, 0.8) rectangle (2.2, 1.2);
  \node[above=5pt, right] at (0.6, 0.8) {q[1]};
  \draw (0, 1.2) rectangle (2.2, 1.6);
  \node[above=5pt, right] at (0.6, 1.2) {q[0]};
  \draw (0, 1.6) rectangle (2.2, 2);
  \node[above=5pt, right] at (0.6, 1.6) {k};
  \draw (0, 2) rectangle (2.2, 2.4);
  \node[above=5pt, right] at (0.6, 2) {p[255]};
  \draw[dotted] (0, 2.4) rectangle (2.2, 2.8);
  \node[above=5pt, right] at (0.7, 2.4) {......};
  \draw (0, 2.8) rectangle (2.2, 3.2);
  \node[above=5pt, right] at (0.4, 2.8) {p[1]};
  \draw (0, 3.2) rectangle (2.2, 3.6);
  \node[above=5pt, right] at (0.3, 3.2) {p[0], \textcolor{blue}{\bf q[255]}};
  \draw[<->, line width=0.8pt](2.4, 3.6) -- (2.4, 2);
  \draw[<->, line width=0.8pt](2.4, 2) -- (2.4, 0);
  \node[left=5pt, right] at (2.6, 2.8) {256 bytes};
  \node[left=5pt, right] at (2.6, 1) {256 bytes};
\end{tikzpicture}

}
\caption{The direct-mapped cache layout (cf.~\cite{ChattopadhyayBRZ17}).}
\label{fig:cache-map}
\vspace{-1ex}
\end{figure}
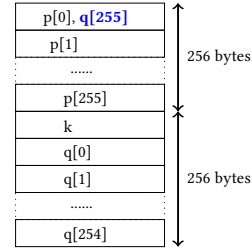

Consider executing the two programs under a 512-byte direct-mapped
cache with one byte per cache line, as shown in
Figure~\ref{fig:cache-map}.  The choice of one-byte-per-cache-line
--- same as in \cite{ChattopadhyayBRZ17} --- is meant to simplify 
analysis without loss of generality.  Specifically, the 256-byte array
\texttt{p} is associated with the first 256 cache lines, while
variable \texttt{k} is associated with the 257-th cache line.  Due to
the finite cache size, \texttt{q[255]} has to share the cache line
with \texttt{p[0]}.

There are two program paths in $P$, each with three memory accesses:
\texttt{load} (line 5), \texttt{load} (line 7 or line 9), and
\texttt{store} (line 11).  However, depending on the value of
\texttt{k}, these three memory accesses may exhibit different cache
behaviors, thus causing data-dependent timing variance.

Assume that \texttt{k}'s value is 0, executing $P$ means taking the
\texttt{then} branch and accessing \texttt{p[0]}, \texttt{q[255]}, and
\texttt{p[0]}.  The first access to \texttt{p[0]} is a cold miss since
the cache is empty at the moment.  The access to
\texttt{q[255]} is a conflict miss because the cache line (shared by
\texttt{q[255]} and \texttt{p[0]}) is occupied by \texttt{p[0]}; as a
result \texttt{q[255]} evicts \texttt{p[0]}.  The next access to
\texttt{p[0]} is also a conflict miss since the cache line is occupied
by \texttt{q[255]}.  All in all, the cache behavior is
\texttt{<miss,miss,miss>} for \texttt{k=0}.

This sequence is also unique in that all other values of \texttt{k}
would produce \texttt{<miss,miss,hit>} as shown at the bottom of
Figure~\ref{fig:motiv-seq}(a).
This means $P$, when running alone, leaks information about
\texttt{k}.  For example, upon observing the delay caused by
\texttt{<miss,miss,miss>} via monitoring, an adversary may infer that
\texttt{k}'s value is \texttt{0}.

Program $P_r$ is a repaired version~\cite{ChattopadhyayBRZ17} where the
\texttt{load} is moved from line 5 to line 9 as in
Figure~\ref{fig:motiv-seq}(b).  Thus, the \texttt{load} accessing
\texttt{p[k]} at line 9 always generates a cold miss
(\texttt{0<k$\leq$255}) or a conflict miss (\texttt{k=0}).
Consequently, the \texttt{store} at line 11 is always a hit.
Thus, for all values of \texttt{k}, the cache behavior remains
\texttt{<miss,miss,hit>} -- no information of \texttt{k} is leaked.

\subsection{New Leak Induced by Concurrency}
\label{sec:concurrent-example}

Although $P_r$ is a valid repair when the program is executed
sequentially, the situation changes when it is executed concurrently
with other threads.
Specifically, if we use one thread ($T_1$) to execute $P_r$ while
allowing a second thread ($T_2$) to run concurrently, $P_r$ may exhibit
new timing leaks.

\begin{figure}
\centering
\framebox[.46\linewidth]{     
  \begin{minipage}{0.45\linewidth}
    {\scriptsize \tt
      \begin{tabbing}
	xxx \= xxx \= \kill
	\textcolor{dkgreen}{\texttt{/*}}\>\textcolor{dkgreen}{\texttt{[Thread T1]\ */}}    \\
	1: \> char p[256];  	 	\\              
	2: \> unsigned char k;  	\\              
	3: \> char q[256];         	\\  
  	4: \>          	\\  
	5: \>  if (k <= 127) 		\\
	6: \>  	~ load reg2, q[255-k] \\
    	7: \>  else \\
    	8: \>  	~ load reg2, q[k-128] \\
    	9: \>  \textbf{load reg1, p[k]} 		\\
    	10: \> add reg1, reg2 \\
    	11: \> store reg1, p[k]
      \end{tabbing}
    }
  \end{minipage}
}
\hspace{.01\linewidth}
\framebox[.46\linewidth]{     
  \begin{minipage}{0.45\linewidth}
    {\scriptsize \tt
      \begin{tabbing}
	xxx \= xxx \= \kill
  	\textcolor{dkgreen}{\texttt{/*}}\>\textcolor{dkgreen}{\texttt{[Thread T2]\ */}}    \\
    	12: \> unsigned char tmp;  	 	\\              
    	13: \> $\textbf{load reg3, tmp}$\\
        14: \> ... \\
        \> \\
        \> \\
        \> \\
        \> \\
        \> \\
        \> \\
        \> \\
        \>     
      \end{tabbing}
    }
  \end{minipage}
}

\vspace{-1ex}
\caption{Concurrent program with side-channel leak.}
\label{fig:motiv-conc}
\vspace{-2ex}
\end{figure}

Figure~\ref{fig:motiv-conc} shows a two-threaded program comprising
$T_1$ and an adversarial $T_2$ that accesses a new variable
\texttt{tmp}.  Assume \texttt{tmp} is mapped to the same cache line as
\texttt{p[1]}.  Then, it is possible for $T_2$ to cause $T_1$ to leak
information of its secret data.
There are various ways of mapping \texttt{tmp} to the same cache line
as \texttt{p[1]}, e.g., by dynamically allocating the memory used by 
\texttt{tmp} or invoking a recursive (or non-recursive)
function within which \texttt{tmp} is defined as a stack variable.

Table~\ref{tbl:interleavings} shows the six interleavings of threads
$T_1$ and $T_2$.  The left half of this table contains three
interleavings where $T_1$ took the \texttt{then} branch of the
if-statement, while the right half contains three interleavings where
$T_1$ took the \texttt{else} branch. In each case, the four columns
show the ID, the execution order, the cache sequence of thread $T_1$,
and the value range of \texttt{k}.
For example, in \texttt{6-9-11-13}, the \texttt{store} at line 11 is a
cache hit because its immediate predecessor (line 9) already loads
\texttt{p[k]} into the cache.  Since the last \texttt{load} at line 13
comes from thread $T_2$, the cache behavior sequence of $T_1$ is
\texttt{<miss,miss,hit>}, denoted \texttt{<m,m,h>} for brevity.

\begin{table}[htb!]
  \caption{Interleavings and thread $T_1$'s cache sequences.}
  \label{tbl:interleavings}
  \centering
  \resizebox{.99\linewidth}{!}{
    \addtolength{\tabcolsep}{-2pt}
    \begin{tabular}{llcc||llcc}
      \toprule
      ID &  Interleaving &Cache-seq &\texttt{k} &ID & Interleaving & Cache-seq &\texttt{k} \\ 	
      \midrule
      1 &{\tt 6-13-9-11} &<m,m,h> &[0,127] &4 &{\tt 8-13-9-11}  &<m,m,h> & (127,255] \\
      2 &{\tt 6-9-11-13} &<m,m,h> &[0,127] &5 &{\tt 8-11-9-13}  &<m,m,h> & (127,255] \\
      3 &{\tt 6-9-13-11} &<m,m,h> &[0,1)$\cup$(1,127] 
                                           &6 &{\tt 8-9-13-11 } &<m,m,h> & (127,255] \\
        &	         &<m,m,m> &1       &  &                 &        &           \\
      \bottomrule
    \end{tabular}
    }
\end{table}

Although context switches between the threads $T_1$ and $T_2$ may occur at any
time in practice, for the purpose of analyzing cache timing leaks, we
assume they occur only before the \texttt{load} and \texttt{store}
statements.  Furthermore, we only focus on these memory accesses when
they are mapped to the same cache line, e.g., between the
\texttt{load} in $T_2$ and statements that access \texttt{p[k]} in
$T_1$.

\begin{figure}
\vspace{1ex}
  \centering
  \scalebox{0.9}{
  \begin{tikzpicture}[fill=blue!20,font=\footnotesize] 
    \draw (0, 0) rectangle (2.2, 0.4);
    \node[above=5pt, right] at (0.6, 0) {q[254]};
    \draw[dotted] (0, 0.4) rectangle (2.2, 0.8);
    \node[above=5pt, right] at (0.7, 0.4) {......} ;	
    \draw (0, 0.8) rectangle (2.2, 1.2);
    \node[above=5pt, right] at (0.6, 0.8) {q[1]};
    \draw (0, 1.2) rectangle (2.2, 1.6);
    \node[above=5pt, right] at (0.6, 1.2) {q[0]};
    \draw (0, 1.6) rectangle (2.2, 2);
    \node[above=5pt, right] at (0.8, 1.6) {k};
    \draw (0, 2) rectangle (2.2, 2.4);
    \node[above=5pt, right] at (0.6, 2) {p[255]};
    \draw[dotted] (0, 2.4) rectangle (2.2, 2.8);
    \node[above=5pt, right] at (0.7, 2.4) {......};
    \draw (0, 2.8) rectangle (2.2, 3.2);
    \node[above=5pt, right] at (0.4, 2.8) {p[1]};
    \node[above=5pt, right, color=blue] at (1, 2.8) {, tmp};
    \draw (0, 3.2) rectangle (2.2, 3.6);
    \node[above=5pt, right] at (0.3, 3.2) {p[0], q[255]};
    \draw (4, 0.0) -- (4, 0.5);
    \draw [orange](3.9, 0.5) rectangle (4.1, 1.0);
    \node [left=22pt, right] at (4.0, 0.7) {p[k]};
    \draw (4, 1.0) -- (4, 1.4);
    \draw [blue](3.9, 1.4) rectangle (4.1, 1.9);
    \node [left=22pt, right] at (4.0, 1.6) {p[k]};
    \draw (4, 1.9) -- (4, 2.3);
    \draw [blue](3.9, 2.3) rectangle (4.1, 2.8);
    \node [left=39pt, right] at (4, 2.5) {q[255-k]};
    \draw (4, 2.8) -- (4, 3.2);
    \node  at (4, 3.4) {\texttt{T1}};
    \draw (5, 0.0) -- (5, 1.0);
    \draw [blue](4.9, 1.0) rectangle (5.1, 1.5);
    \node [left, right=5pt] at (5.1, 1.2) {tmp};
    \draw (5, 1.5) -- (5, 3.2);
    \node  at (5, 3.4) {\texttt{T2}};
    \draw [dashed,thick,red](4.3, 3.2)--(4.3, 1.3)--(5.25,1.8);
    \draw [dashed,thick,red](5.25,1.8)--(5.25,0.8)--(4.25,1.1);
    \draw [->,dashed,thick,red](4.3, 1.1)--(4.3, 0.0);
  \end{tikzpicture}

}
  \caption{Interleaving \texttt{6-9-13-11} and the cache layout.}
  \label{fig:specific_interleaving}
\vspace{-2ex}
\end{figure}
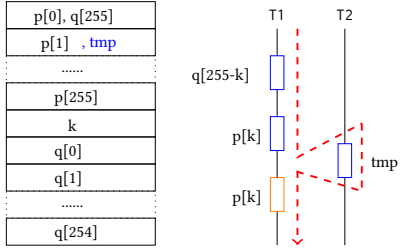

We use Figure~\ref{fig:specific_interleaving} to show details of
\texttt{6-9-13-11}.  The blue and orange rectangles represent the
\texttt{load} and \texttt{store} accesses, respectively, and the red
dashed poly-line shows their execution order.
The first three \texttt{load} operations all cause cache misses,
whereas the last \texttt{store} could be a cache hit if (\texttt{k!=1})
and a cache miss if (\texttt{k=1}).
When ($\texttt{k=1}$), the four memory accesses would be
$\texttt{q[254]}$, $\texttt{p[1]}$, $\texttt{tmp}$, and
$\texttt{p[1]}$.  The first two trigger cold misses.  The third
one ($\texttt{tmp}$) triggers a conflict miss as the cache line was
occupied by $\texttt{p[1]}$.  Evicting this cache line would then lead
to another conflict miss for the subsequent $\texttt{store}$ to
$\texttt{p[1]}$.

The examples presented so far show that, even for a timing-leak-free
program ($T_1$), running it concurrently with another thread ($T_2$)
may cause it to exhibit new timing leaks.  This is the case even if
the two threads ($T_1$ and $T_2$) are logically independent of each
other.  In other words, they do not need to share variables or
communicate through messages; they can affect each other's timing
behaviors by sharing the same cache system.

\subsection{Adversarial Symbolic Execution}
\label{sec:app-scenarios}

The goal of developing a new symbolic execution method is to detect
such timing leaks.  More specifically, we are concerned with two
application scenarios for \SymSC{}, depending on whether the
adversarial thread ($T_2$) exists in the given program or not.

%
\textit{Case 1. Thread $T_2$ is given, together with fixed addresses
  of the memory region accessed by $T_2$.}  In this case, $T_2$ is an
integral part of the concurrent system that also contains the
security-critical computation in $T_1$.  Since the only source of
nondeterminism is thread interleaving, our tool aims to check if the
concurrent system itself has timing leaks.

%
\textit{Case 2. Thread $T_2$ is not given, but created by our tool,
  and thus the addresses of the memory region accessed by $T_2$ are
  assumed to be symbolic.}  This is when, inside the cache layout of
Figure~\ref{fig:specific_interleaving}, the address of \texttt{tmp} would be 
made symbolic, thus allowing it to be mapped to any cache line (as
opposed to be fixed to the 2nd line).  There are now two sources of
non-determinism: thread interleaving and memory layout.  Our tool 
explores both to check if $T_1$ may leak information due
to interference from $T_2$.

%
In the second case, when $T_2$ executes a memory load instruction $t$, for example, the
symbolic address $addr$ may be mapped to any cache line.  The purpose
of having such aggressive adversarial addressing is to allow
\SymSC{} to conduct a (predictive) \emph{what-if} analysis: it
searches all potential memory layouts to check if there exists one
that allows $T_2$ to cause a timing leak.

\section{The Threat Model}
\label{sec:prelim}

We now review the technical background and present the threat model,
which defines what an adversary can or cannot do.

\subsection{Cache and the Timing Side Channels}
\label{sec:cache}

The execution time of a program depends on the CPU cycles taken to
execute the instructions and the time needed to access memory.
The first component is easy to compute but also less
important in practice, because security-critical applications often execute the
same set of instructions regardless of values of their sensitive variables~\cite{WuGSW18}. 
In contrast, leaks are more likely to occur in the second component:
the time taken to access memory.  Compared to the time needed to
execute an instruction, which may be 1-3 clock cycles, the time taken
to access memory, during a cache miss, may be tens or even
hundreds of clock cycles.

There are different types of cache based on the size, associativity
and replacement policy.  For ease of comprehension, we use
direct-mapped cache with LRU policy in this paper, but other cache
types may be handled similarly.  Indeed, during our
experiments, both direct-mapped cache and 4-way set-associative cache
were evaluated and they led to similar analysis results.


We assume the security-critical program $P$ implements a function $c
\leftarrow f(k,x)$, where $k$ is the sensitive input (secret), $x$ is
the non-sensitive input (public), and $c$ is the output.  In block
ciphers, for example, $k$ would be the cryptographic key, $x$ would be
the plaintext, $c$ would be the ciphertext, and $f$ would be the
encryption or decryption procedure.

Let the execution time of $P$ be $\tau_P(k,x)$.  Since there may be
multiple paths inside $P$, when referring to a particular path $p\in
P$, we use $\tau_p(k,x)$.  But if there is no ambiguity, we may omit
the detail and simply use $\tau(k,x)$.
We say P is leak-free if $\tau(k,x)$ remains the same for all input
values.  That is,
\[ \forall x, k_1,k_2 ~.~  \tau(k_1,x) = \tau(k_2,x) \]
Here $k_1$ and $k_2$ are two arbitrary values of $k$.  Since in
practice, decision procedures (e.g., SMT solvers) are designed for
checking satisfiability, instead of proving the validity of a 
formula, we try to falsify it by checking the formula below:
\[  \exists x, k_1,k_2 ~.~  \tau(k_1,x) \neq \tau(k_2,x) \]
Here, we search for two values of $k$ that can lead to differences.

If the set of instructions executed by $P$ remains the same, we only
need to check whether $\tau(k_1,x)$ and $\tau(k_2,x)$ have the same number
of cache hits and misses.
Furthermore, in our threat model where the attacker can only observe
(passively) the execution time of $P$, but not control or observe
$x$, we can reduce the computational cost by fixing a value
$\overline{x}$ of $x$ arbitrarily and then checking if $\tau(k_1)$ and
$\tau(k_2)$ have the same number of cache hits and misses.

\subsection{Example of an Attack}
\label{sec:hpn-ssh}

Now, we show a concrete example of exploiting cache timing leaks in
concurrent systems.  The goal is to illustrate what an adversary may
be able to achieve in practice.

\begin{figure}
\vspace{1ex}
\centering

{\scriptsize \tt uint8\_t *buf = 0;}
{\scriptsize \tt uint32\_t size = INPUT\_SIZE;}
{\scriptsize \tt uint32\_t idx = 0;}
\vspace{.5ex}

\framebox[.46\linewidth]{     
  \begin{minipage}{0.45\linewidth}
    {\scriptsize \tt
      \begin{tabbing}
	xxx \= xxx \= \kill
	\textcolor{dkgreen}{\texttt{/*}}\>\textcolor{dkgreen}{\texttt{[Thread T1]\ */}}    \\
	1: \> uint8\_t S[256] = \{0x4b,...\};  	 	\\              
	2: \> uint8\_t out[64] = \{0\};         	\\  
	3: \> for(int i=0; i<size; )\\
	4: \> ~~if (i < idx) 		\\
	5: \> ~~~~memcpy(out,buf+i,64); \\
	6: \> ~~~~for (int j=0;j<64;j++,i++) \\
	7: \> ~~~~~~out[j] \&= S[key[j]];\\
	8: \> ~~~~write(out, ...);\\
    9: \> ~~else \\
    10:\> ~~~~sleep(50);
      \end{tabbing}
    }
  \end{minipage}
}
\hspace{.01\linewidth}
\framebox[.46\linewidth]{     
  \begin{minipage}{0.45\linewidth}
    {\scriptsize \tt
      \begin{tabbing}
	xxx \= xxx \= \kill
  	\textcolor{dkgreen}{\texttt{/*}}\>\textcolor{dkgreen}{\texttt{[Thread T2]\ */}}    \\
		11: \> ......\\    	
    	12: \> buf=(uint8\_t *)malloc(size);\\
    	13: \> while(idx<size)\\
        14: \> ~~memcpy(buf+idx,read(...),64); \\
        15: \> ~~idx+=64;\\
        16: \> ......\\
        \> \\
        \> \\
        \> \\
      \end{tabbing}
    }
  \end{minipage}  
} 

\vspace{1ex}
\scalebox{.82}{
\begin{tikzpicture}[fill=blue!20,font=\footnotesize]   
  \draw (0.4, 0.4) rectangle (4.4, 0.8);
  \node[above=5pt, right] at (2.0, 0.4) {......} ;	
  \draw (0.4, 0.8) rectangle (4.4, 1.2);
  \node[above=5pt, right, color=blue] at (0.5, 0.8) {S[192]-S[255]};
  \draw (0.4, 1.2) rectangle (4.4, 1.6);
  \node[above=5pt, right, color=blue] at (0.5, 1.2) {S[128]-S[191]};
  \draw (0.4, 1.6) rectangle (4.4, 2);
  \node[above=5pt, right, color=blue] at (0.5, 1.6) {S[64]-S[127]};
  \draw [fill=yellow] (0.4, 2) rectangle(4.4, 2.4);
  \node[above=5pt, right, color=blue] at (0.5, 2) {S[0]-S[63]  ,};
  \node[above=5pt, right] at (2.2, 2) {buf[960]-buf[1023]};
  \draw (0.4, 2.4) rectangle (4.4, 2.8);
  \node[above=5pt, right] at (2.0, 2.4) {......};
  \draw (0.4, 2.8) rectangle (4.4, 3.2);
  \node[above=5pt, right] at (2.2, 2.8) {buf[0]-buf[63]};
  \draw (0.4, 3.2) rectangle (4.4, 3.6);
  \node[above=5pt, right] at (2, 3.2) {......};
  
  \draw (0.4, 3.65) -- (0.4, 3.75);
  \draw [->] (0.4, 3.7) -- (2, 3.7);
  \draw (4.4, 3.65) -- (4.4, 3.75);
  \draw [->] (4.4, 3.7) -- (3.0, 3.7);
  \node at (2.5,3.7) {64 bytes};
  
  \draw(4.5, 3.6) -- (4.6,3.6);
  \draw [->] (4.55, 3.6) -- (4.55, 2.2);
  \draw [->] (4.55, 0.4) -- (4.55, 1.8);
  \draw(4.5, 0.4) -- (4.6,0.4);
  \node at (4.7,2.0) {32KB};
  
\end{tikzpicture}}
\hspace{.01\linewidth}
\scalebox{.86}{
\begin{tikzpicture}[fill=blue!20,font=\footnotesize] 
	\draw (5.6, 0.0) -- (5.6, 0.5);
	\draw (5.6, 0.5) .. controls (5.85, 0.6) .. (5.9, 1.0);
	\draw [blue](5.8, 1.0) rectangle node{7} (6.0, 1.3);
	\draw [->](5.87, 0.8) .. controls (6.5, 1.3) .. (5.9, 1.4);
	\node  at (6.2, 0.8) {\texttt{j<64}};
	
	\draw (5.9, 1.3) -- (5.9, 1.7);
	\draw [blue](5.8, 1.7) rectangle node{5} (6.0, 2.0);
	\draw (5.9, 2.0) .. controls (5.825, 2.2) .. (5.6, 2.3);
	\node  at (6.2, 2.2) {\texttt{i<idx}};
	
	\draw (5.6, 0.5) .. controls (5.3, 0.6) .. (5.3, 1.3);
	\draw [blue](5.16, 1.3) rectangle node{10} (5.44, 1.6);
	\draw (5.3, 1.6) .. controls (5.4, 2.2) .. (5.6, 2.3);

	\draw [->](5.6, 0.3) .. controls (4.7, 0.4) and (4.7, 2.2) .. (5.6, 2.5);
	\node  at (5.1, 2.55) {\texttt{i<size}};

    \draw (5.6, 2.3) -- (5.6, 2.9);
    \node  at (5.6, 3.0) {\texttt{T1}};

    \draw (7.6, 0.0) -- (7.6, 1.0);
    \draw (7.46, 1.0) rectangle node{14} (7.74, 1.3);
    \draw (7.6, 1.3) -- (7.6, 1.6);
    \draw (7.46, 1.6) rectangle node{13} (7.74, 1.9);
	\draw [->](7.6, 0.7) .. controls (9.1, 1.5) and (8.1,2.0) .. (7.6, 2.2);
    \node  at (8.4, 2.2) {\texttt{idx<size}};
    \draw (7.6, 1.9) -- (7.6, 2.9);
    \node  at (7.6, 3.0) {\texttt{T2}};
    
    \draw [->,dotted,thick,red](7.55, 2.05) .. controls (8.5, 1.5) and (8.1,1.5) .. (5.9, 1.5);
    \node [color=red] at (6.75, 1.7) {idx: 960};
    
  \end{tikzpicture}

}

\vspace{-2ex}
\caption{A two-threaded encryption program.}
\label{fig:motiv-conc-new}
\vspace{-2ex}
\end{figure}
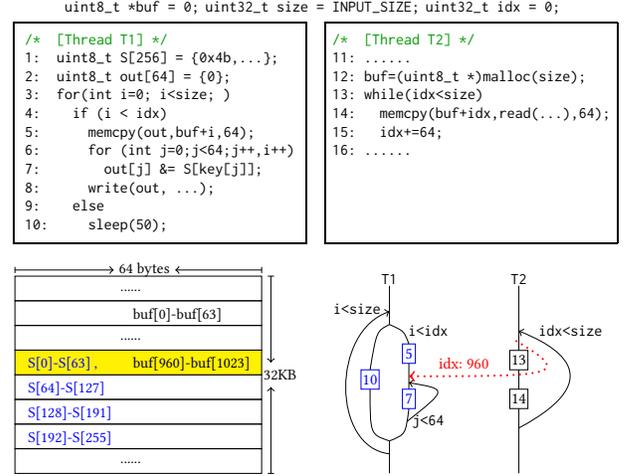

Figure~\ref{fig:motiv-conc-new} shows a two-threaded program, its
cache mapping, and the thread-local control flows. Initially,
\texttt{T2} allocates a memory area (\texttt{buf}) whose size matches
the input.  Although the input size may be arbitrary, here, we assume
it is an integral multiple of 64, e.g., 1024 bytes
(\texttt{INPUT\_SIZE}=1024). In the \textit{while}-loop (line 14)
\texttt{T2} reads 64 bytes from input every time to fill
\texttt{buf}. Thread \texttt{T1} tracks the progress (\texttt{idx}) of
\texttt{T2} (line 4) and repeatedly retrieves 64-byte data from
\texttt{buf} to the array \texttt{out} (line 5). The encryption on
\texttt{out} involves the S-Box array \texttt{S} and a given
\texttt{key} (lines 6-7). Once the data is encrypted, \texttt{T1}
sends it out (line 8). When \texttt{T1} finds that \texttt{buf} runs
out of data, it sleeps for 50ms (line 10).

First, we explain why the program has a timing leak. We use a 32KB
direct-mapped cache here and set each cache line to 64 bytes. The 
S-Box array \texttt{S} hence maps to 4 cache lines and the \texttt{buf}
array  maps to 16 cache lines. For brevity, we only focus on the
important arrays (\texttt{S} and \texttt{buf}) while assuming 
other variables do not affect the cache mapping. Furthermore, we 
assume  \texttt{S} and \texttt{buf} share one cache line as 
highlighted in Figure~\ref{fig:motiv-conc-new}.

The graph in Figure~\ref{fig:motiv-conc-new} shows an interleaving 
of \texttt{T1} and \texttt{T2}, where the dotted red arrow represents a
context switch after \texttt{T2} executes the \texttt{memcpy} statement
(line 14) while \texttt{T1} just reaches the \textit{for}-loop at line 6. 
The text above the arrow means \texttt{idx}'s value is 960 at the moment, 
indicating thread \texttt{T2} has just accessed the last 64 bytes of
\texttt{buf} at line 14.

After the context switch, \texttt{T1} enters the \textit{for}-loop (line 6) 
and reads \texttt{S[key[j]]} at line 7. Note that the offset to \texttt{S}'s 
base address depends on \texttt{key[j]}, thus different keys may make thread 
\texttt{T1} access different items of \texttt{S}. We pick two 64-byte keys 
$\mathbf{k1}$ and $\mathbf{k2}$ which differ in the first eight bits: 
10000000 for $\mathbf{k1}$ and 00000000 for $\mathbf{k2}$. Using $\mathbf{k1}$, 
thread \texttt{T1} first reads \texttt{key[0]} and \texttt{S[128]}. The 
access to \texttt{S[128]} would lead to a cache hit if \texttt{i} is greater than 
63. This is because after the \textit{for}-loop (lines 6-7) finishes once 
(\texttt{i}=64), \texttt{S[128]} is already mapped to cache and no further 
accesses evict it.

In contrast, with $\mathbf{k2}$, thread \texttt{T1} loads \texttt{S[0]} which maps 
to the cache line shared with \texttt{buf[960-1023]}. Recall that, before the 
context switch, \texttt{T2} just accessed the area starting from \texttt{buf+idx} 
(\texttt{buf[960]}). Consequently \texttt{T1}'s access to \texttt{S[0]} causes a 
conflict miss because the shared cache line was occupied by \texttt{buf}. Thus, 
we find a leak: two keys ($\mathbf{k1}$ and $\mathbf{k2}$) leading to divergent 
cache behaviors at a program location due to thread interleaving.

Next, we discuss how this leak may be exploited. The 
leak is due to the sharing of cache between \texttt{S} and
\texttt{buf}, which is crucial to our threat model. In this program,
\texttt{S} has a fixed size while \texttt{buf} is dynamically
allocated at run time based on the input data. Furthermore,
\texttt{INPUT\_SIZE} is a variable affected by the external input.
Although the actual input size cannot be arbitrarily large in
practice, for this exploit to work, it only needs to be larger than
the total cache size, which is 32KB.

Thus, the attacker could mutate the input to alter the buffer
size, hence affecting the memory layout. Furthermore, real
applications sometimes use relatively large fixed buffers.  For
example, in OpenSSH~\cite{openssh}, the \texttt{scp} program has a
16KB buffer for \texttt{COPY\_BUFLEN} and the \texttt{sftp} program
has a 32KB buffer for \texttt{DEFAULT\_COPY\_BUFFER}. Moreover,
OpenSSH's \texttt{SSHBUF\_SIZE\_MAX} buffer for a socket channel is as
large as 256MB. These large buffers allow room for attackers to
construct the desired cache layout.

\ignore{E.g., OpenSSH~\cite{openssh}'s SCP has a 16KB \texttt{COPY\_BUFLEN}, which 
is a half of a modern processor's 32KB L1 cache. The sftp program has a even larger
\texttt{DEFAULT\_COPY\_BUFFER} of 32KB. Moreover, in the socket channel OpenSSH has
\texttt{SSHBUF\_SIZE\_MAX} of 256MB.}

We have found a similar scenario in the open-source implementation of
HPN-SSH~ \cite{hpn-ssh}, which is an enhancement of
OpenSSH~\cite{openssh} by leveraging multi-threading to accelerate the
data encryption.
Figure~\ref{fig:hpn-ssh} shows the code snippet directly taken from
the HPN-SSH~\cite{hpn-ssh} repository: On the left-hand side are
threads created to run the \texttt{thread\_loop} function, shown on
the right-hand side, which repeatedly calls \texttt{AES\_encrypt} to
encrypt data given by the user (line 327).  By controlling the size
and content of the data, as well as the number of threads, a malicious
user is able to affect both the memory layout and the thread
interleaving.

\begin{figure}
\vspace{1ex}
\centering
\framebox[.46\linewidth]{     
  \begin{minipage}{0.45\linewidth}
    {\scriptsize \tt
      \begin{tabbing}
	xxx \= xxx \= \kill
	\textcolor{dkgreen}{\texttt{/*}}\>\textcolor{dkgreen}{\texttt{cipher-ctr-mt.c\ */}}    \\
	\> ...  	 	\\              
	504: \> for(i=0;i<CIPHER\_THREADS;i++)\{        	\\  
	\> ~~......\\
	507: \> ~~pthread\_create(...,\\
	\> ~~~~~~~~~~~~thread\_loop,...); \\
	\> ~...... \\
    509: \>\} 
      \end{tabbing}
    }
  \end{minipage}
}
\hspace{.01\linewidth}
\framebox[.46\linewidth]{     
  \begin{minipage}{0.45\linewidth}
    {\scriptsize \tt
      \begin{tabbing}
	xxx \= xxx \= \kill
  	\textcolor{dkgreen}{\texttt{/*}}\>\textcolor{dkgreen}{\texttt{cipher-ctr-mt.c\ */}}    \\
    	238: \> static void* \\
    	239: \> thread\_loop(void *x) \{~~~~~~~~ \\
    	\> ~~......\\
        326: \> ~~for(i=0;i<KQLEN;i++) \{ \\
        327: \> ~~~~AES\_encrypt(q->ctr,\\
        \> ~~~~~~~~~~q->keys[i],\&key);\\
        \> ~~......
      \end{tabbing}
    }
  \end{minipage}  
}

\vspace{-2ex}
\caption{Concurrency-related code in HPN-SSH~\cite{hpn-ssh}.}
\label{fig:hpn-ssh}
\vspace{-2ex}
\end{figure}

In our experimental evaluation (Section~\ref{sec:evaluation}), we will
show that the AES subroutine from OpenSSL indeed has cache timing
leaks, which may subject HPN-SSH to attack scenarios similar to the
one illustrated in Figure~\ref{fig:motiv-conc-new}.

\section{Adversarial Symbolic Execution}
\label{sec:adversarialSE}

We first present the baseline algorithm for concurrent programs.
Then, we enhance it to search for cache timing leaks.

\subsection{The Baseline Algorithm}
\label{sec:baseline}

Following Guo et al.~\cite{GuoKWYG15}, we assume the entire program
consists of a finite set $\{T_1,\ldots,T_n\}$ of threads where each
thread $T_i$ ($1\leq i\leq n$) is a sequential program.  Without loss
of generality, we assume $T_1$ is critical and any of $T_2,\ldots,T_n$
may be adversarial.
Let $st$ be an instruction in a thread.  Let \emph{event} $e=\langle
tid, l, st, l' \rangle$ be an instance of $st$, where $l$ and $l'$ are
thread-local locations before and after executing $st$.  A
\emph{global location} is a tuple $s = \langle
l_1,\ldots,l_n\rangle$ where each $l_i$ is a location in $T_i$.
Depending on the type of $st$, an event may have one of the following types:
\begin{itemize*}
  \item $\alpha$-event, which is an assignment $v_l := \mathit{exp}_l$
    where $v_l$ is a local variable and $exp_l$ is an expression in
    local variables.
  \item $\beta$-event, which is a local branch denoted
    \textcode{assume$(\mathit{cond}_l)$} where the condition $cond_l$ is
    expressed in local variables.
  \item $\gamma$-event, which is a \texttt{load} from global memory of
    the form $v_l := v_g$, a \texttt{store} to global memory of the
    form $v_g :=\mathit{exp}_l$, or a thread synchronization
    operation.
\end{itemize*}
For an \textcode{if(c)-else} statement, we use \textcode{assume$(c)$}
to denote the then-branch, and \textcode{assume($\neg c$)} to denote
the else-branch.  Since $c$ is expressed in local variables or local
copies of global variables, $\beta$-events are local branching points
whereas $\gamma$-events are thread interleaving points. Both $\beta$-
and $\gamma$-events contribute to the state-space explosion problem.
In contrast, $\alpha$-events are local to their own threads.
Details on handling of language features such as pointers and function
calls are omitted, since they are orthogonal issues addressed by
existing symbolic execution tools~\cite{CadarDE08,CiorteaZBCC09}.


Algorithm~\ref{alg:baseline} shows the baseline symbolic execution
procedure that follows the prior
work~\cite{CiorteaZBCC09,BerganGC14,GuoKWYG15} except that, for the
purpose of detecting timing leaks, it considers two events as
\emph{dependent} also when they are mapped to the same cache line.
Here, an execution is characterized by
$\pi=(\mathit{in},\mathit{sch})$ where $\mathit{in}=\{k,x\}$ is the
data input and $\mathit{sch}$ is the thread schedule, corresponding to
a total order of events $e_1\ldots e_n$, and $\emph{Stack}$ is a
container for symbolic states.  Each $s\in\emph{Stack}$ is a tuple
$\langle \mem, \pcon, \mathit{branch},\mathit{enabled}, crt \rangle$,
where $\mem$ is the symbolic memory, $\pcon$ is the path condition,
$\mathit{branch}$ is the set of branching ($\beta$) events,
$\mathit{enabled}$ is the set of thread interleaving ($\gamma$)
events, and $\mathit{crt}$ is the event chosen to execute at $s$.

\begin{algorithm}
\caption{Baseline Symbolic Execution Procedure.}
\label{alg:baseline}
{\footnotesize
\SetAlgoLined
\setstretch{0.5}
\DontPrintSemicolon
\nonl \textbf{Initially}: State stack $\emph{Stack}$ = $\emptyset$;\\
\nonl Start \textbf{\SymSC{}($s_0$)} with the initial symbolic state $s_0$.\\
\textbf{\SymSC{}}(State $s$)\\
\Begin{
	$\emph{Stack}$.push($s$);\;
	\uIf{$s$ is a thread-local branching point}{
		\For{$t \in s.branch$ {\bf and} $s$.$\pcon\wedge t$ is satisfiable}{
			\textbf{\SymSC{}}($\mathit{NextSymbolicState}$($s$, $t$));~~ \textcolor{gray}{// $\beta$ event}
		}
	}
	\uElseIf{$s$ is a thread interleaving point}{
		\For{$t \in s.enabled$}{
			\textbf{\SymSC{}}($\mathit{NextSymbolicState}$($s$, $t$));~~ \textcolor{gray}{// $\gamma$ event (enhanced)}
		}
	}\uElseIf{$s$ is other sequential computation}{
			\textbf{\SymSC{}}($\mathit{NextSymbolicState}$($s$, $s.crt$));~~~~~~~~~ \textcolor{gray}{// $\alpha$ event}
	}\Else{
               terminate at $s$;\;
        }
	$\emph{Stack}$.pop();\;
}
\BlankLine
$\mathit{NextSymbolicState}$(State $s$, Event $t$)\;
\Begin{
	$s.crt \leftarrow t$;\;
	$s'\leftarrow$ Execute the event $t$ in the state $s$;\;
	\textbf{return} $s'$;
}
}
\end{algorithm}

At the beginning, the stack is empty and the entry is the initial
state $s_0$.  Then, depending on the type of the state $s$, we may
execute a local branch (line 4), perform a context switch (line 8), or
execute a sequential computation (line 12).  In all cases, \SymSC{} is
invoked again on the new state.

Sub-procedure \emph{NextSymbolicState} takes the current state $s$ and
to-be-executed event $t$ as input, and returns the new state $s'$ as
output: $s'$ is the result of executing $t$ at $s$.  We omit details
since they are consistent with existing symbolic execution methods~\cite{GuoKWYG15,GuoKW16,GuoWW17,YiYGWLZ18,YuZW17}.

Also note that, in the prior work, symbolic execution would allow
interleavings between global ($\gamma$) events only if they have
\emph{data conflicts}, i.e., they are from different threads,
accessing the same memory location, and at least one of them is a
write.  This is because only such accesses may lead to different
states if they are executed in different orders.
However, in our case, whether these events are mapped to the same
cache line also matters.

\subsection{Enhanced Algorithm}

We enhance the baseline algorithm to arrive at Algorithm~\ref{alg:symsc}, where
the main difference is in the interleaving points.
Upon entering the \emph{for}-loop at line 5, we first check if an
enabled event $t$ may lead to a timing leak by invoking
\emph{DivergentCacheBehavior(s,t)}.
Details of the subroutine will be presented in
Section~\ref{sec:adversarialCM}, but at the high level, it constructs
a cache behavior constraint $\tau_t$ and then searches for two values,
$\overline{k_1}$ and $\overline{k_2}$, such that
$\tau_t(\overline{k_1})\neq\tau_t(\overline{k_2})$.

Since detecting such divergent behaviors is computationally expensive,
prior to invoking the subroutine, we make sure that event $t$ indeed
may be involved in an adversarial interleaving. This is determined by
\emph{AdversarialAccess(s,t)} which checks if (1) $t$ comes from the
critical thread $T_1$ and (2) there exists a previously executed event
$t' = s'.crt$ where $s'\in \mathit{Stack}$ and the two events ($t$ and
$t'$) are mapped to the same cache line.

\begin{algorithm}
\caption{Symbolic Execution in \SymSC{}}
\label{alg:symsc}
{\footnotesize
\SetAlgoLined
\setstretch{0.5}
\DontPrintSemicolon
\nonl \textbf{Initially}: State stack $\emph{Stack}$=$\emptyset$;\\
\nonl Start \textbf{\SymSC{}($s_0$)} with the initial symbolic state $s_0$.\\
\textbf{\SymSC{}}(State $s$)\\
\Begin{
	......\;
	\uElseIf{$s$ is a global-memory access point}{
		\For{$t \in s.\textit{enabled}$  }{
			\uIf{\textcolor{darkblue}{DivergentCacheBehavior($s$, $t$)}}{
				\textcolor{darkblue}{generate test case;}\;
				\textcolor{darkblue}{terminate at $s$;}\;
			}\Else{
				\textbf{\SymSC{}}($\mathit{NextSymbolicState}$($s$, $t$));\;	
			}
		}
	}
	......\;
}
\textcolor{black}{
~~~\\
$\mathit{DivergentCacheBehavior}$(State $s$, Event $t$)\;
\Begin{
    \If{\textcolor{darkblue}{AdversarialAccess($s$, $t$)}}{			
	$\tau_t \leftarrow$ compute $t$'s cache hit constraint;\;
	\uIf{$\exists k,k'$ such that $\tau_t(k)\neq \tau_t(k')$}{
		\textbf{return} $\mathit{true}$;\;
	}
    }
    \textbf{return} $\mathit{false}$;\;
}
}
\textcolor{black}{
~~~\\
$\mathit{AdversarialAccess}$(State $s$, Event $t$)\;
\Begin{	
    \uIf{$t$ is from the critical thread}{
        let $s'\in\emph{Stack}$ and $t' = s'.\textit{crt}$;\;
 	\uIf{$\exists t' ~.~$ $t$ and $t'$ may map to same cache line }{
		\textbf{return} $\mathit{true}$;\;
	}
   }
   \textbf{return} $\mathit{false}$;\;
}
}
}
\end{algorithm}

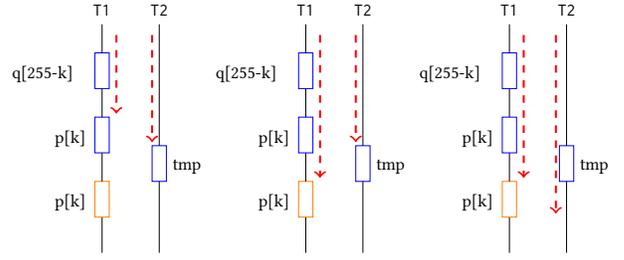
\begin{figure}
\vspace{1ex}
\scalebox{0.95}{
	\begin{tikzpicture}[fill=blue!20,font=\footnotesize]  
    			\draw (1, 0.0) -- (1, 0.5);
    			\draw [orange](0.9, 0.5) rectangle (1.1, 1.0);
    			\node [left=22pt, right] at (1.0, 0.7) {p[k]};
    			\draw (1, 1.0) -- (1, 1.4);
    			\draw [blue](0.9, 1.4) rectangle (1.1, 1.9);
    			\node [left=22pt, right] at (1.0, 1.6) {p[k]};
    			\draw (1, 1.9) -- (1, 2.3);
    			\draw [blue](0.9, 2.3) rectangle (1.1, 2.8);
    			\node [left=39pt, right] at (1.0, 2.5) {q[255-k]};
    			\draw (1, 2.8) -- (1, 3.2);
    			\node at (1, 3.4) {\texttt{T1}};
    			\draw (1.8, 0.0) -- (1.8, 1.0);
    			\draw [blue](1.7, 1.0) rectangle (1.9, 1.5);
    			\node [left, right=5pt] at (1.7, 1.2) {tmp};
    			\draw (1.8, 1.5) -- (1.8, 3.2);
    			\node  at (1.8, 3.4) {\texttt{T2}};
    			\draw [->,dashed,thick,red](1.2, 3.05)--(1.2, 1.95);
    			\draw [->,dashed,thick,red](1.7, 3.05)--(1.7, 1.55);
    			\draw (3.85, 0.0) -- (3.85, 0.5);
    			\draw [orange](3.75, 0.5) rectangle (3.95, 1.0);
    			\node [left=22pt, right] at (3.85, 0.7) {p[k]};
    			\draw (3.85, 1.0) -- (3.85, 1.4);
    			\draw [blue](3.75, 1.4) rectangle (3.95, 1.9);
    			\node [left=22pt, right] at (3.85, 1.6) {p[k]};
    			\draw (3.85, 1.9) -- (3.85, 2.3);
    			\draw [blue](3.75, 2.3) rectangle (3.95, 2.8);
    			\node [left=39pt, right] at (3.85, 2.5) {q[255-k]};
    			\draw (3.85, 2.8) -- (3.85, 3.2);
    			\node  at (3.85, 3.4) {\texttt{T1}};
    			\draw (4.65, 0.0) -- (4.65, 1.0);
    			\draw [blue](4.55, 1.0) rectangle (4.75, 1.5);
    			\node [left, right=5pt] at (4.55, 1.2) {tmp};
    			\draw (4.65, 1.5) -- (4.65, 3.2);
    			\node  at (4.65, 3.4) {\texttt{T2}};
    			\draw [->,dashed,thick,red](4.05, 3.05)--(4.05, 1.05);
    			\draw [->,dashed,thick,red](4.55, 3.05)--(4.55, 1.55);
    			\draw (6.7, 0.0) -- (6.7, 0.5);
    			\draw [orange](6.6, 0.5) rectangle (6.8, 1.0);
    			\node [left=22pt, right] at (6.7, 0.7) {p[k]};
    			\draw (6.7, 1.0) -- (6.7, 1.4);
    			\draw [blue](6.6, 1.4) rectangle (6.8, 1.9);
    			\node [left=22pt, right] at (6.7, 1.6) {p[k]};
    			\draw (6.7, 1.9) -- (6.7, 2.3);
    			\draw [blue](6.6, 2.3) rectangle (6.8, 2.8);
    			\node [left=39pt, right] at (6.7, 2.5) {q[255-k]};
    			\draw (6.7, 2.8) -- (6.7, 3.2);
    			\node  at (6.7, 3.4) {\texttt{T1}};
    			\draw (7.5, 0.0) -- (7.5, 1.0);
    			\draw [blue](7.4, 1.0) rectangle (7.6, 1.5);
    			\node [left, right=5pt] at (7.4, 1.2) {tmp};
    			\draw (7.5, 1.5) -- (7.5, 3.2);
    			\node  at (7.5, 3.4) {\texttt{T2}};
    			\draw [->,dashed,thick,red](6.9, 3.05)--(6.9, 1.05);
    			\draw [->,dashed,thick,red](7.35, 3.05)--(7.35, 0.55);
		\end{tikzpicture}
}

  \caption{The three interleavings generated by \SymSC{}.}
  \label{fig:symsc_interleaving}
\vspace{-2ex}
\end{figure}

%
For our running example in Figure~\ref{fig:motiv-conc}, in particular,
Algorithm~\ref{alg:symsc} would explore the first three interleavings
in Table~\ref{tbl:interleavings} before detecting the leak.  The
process is partially illustrated by
Figure~\ref{fig:symsc_interleaving}, where events $t_1$:\texttt{load
  q[255-k]}, $t_2$:\texttt{load p[k]} and $t_3$:\texttt{store p[k]}
belong to thread $T_1$ whereas $t_4$:\texttt{load tmp} belongs to
thread $T_2$.

Assume $T_1$ executes $t_1$ to reach $t_2$ and $T_2$ is about to
execute $t_4$: this corresponds to the figure on the left.  At this
moment, \emph{s.enabled} = \{ $t_2$, $t_4$ \}.  If $t_4$ is executed
before $t_2$, \emph{AdversarialAccess($s$,$t_2$)} would evaluate to
true because $t_2$ comes from the critical thread and \texttt{p[k]}
may be mapped to the same cache line as \texttt{tmp} accessed by
$t_4$.  However, there is no timing leak at $t_2$, because
\texttt{p[k]} differs from $t_1$'s access \texttt{q[255-k]}, meaning
the cache behavior at $t_2$ remains the same for all values of $k$.

If $t_2$ were executed before $t_4$, we would have the second scenario
in Figure~\ref{fig:symsc_interleaving}.  At this moment,
\emph{s.enabled} = \{ $t_3, t_4$ \}.  If $t_4$ is executed after
$t_3$, the interleaving would be \texttt{6-9-11-13}, which does not
have timing leaks either.
But if $t_4$ were executed before $t_3$, we would have the third
scenario in Figure~\ref{fig:symsc_interleaving}, where
\emph{AdversarialAccess($s,t_3$)} evaluates to true, $\tau_{t_3}(k)$
evaluates to \emph{false} for (\texttt{k}=1) but to \emph{true} for
(\texttt{k}$\neq$1)$\wedge$(\texttt{k}$\leq$127), as shown in
Table~\ref{tbl:individual_constraint}, leading to divergent cache
behaviors in \texttt{6-9-13-11}.

\section{Adversarial Cache Analysis}
\label{sec:adversarialCM}

Our method for detecting divergent cache behaviors is as follows.
First, it constructs the behavioral constraint for each memory access.
Then, it solves the constraint to compute a pair of sensitive values
that allow the constraint to return divergent results.

\subsection{Cache Modeling}
\label{sec:model-cache}

Recall that the entire program contains $T_1$ and $T_2$, among other
threads, where $T_1$ invokes the critical computation and $T_2$ is
potentially adversarial.  During symbolic execution, $\SymSC$ conducts
context switches when $\texttt{load}$ or $\texttt{store}$ instructions
may be mapped to the same cache line.  Here, each interleaving $p$
corresponds to a data input $in$ and a thread schedule $sch$.  The
data input is divided further into $in = \{k,x\}$, where $k$ is
sensitive (secret) and $x$ is non-sensitive (public).  Whenever the
value of $x$ is immaterial, we assume $in=\{k\}$.
\begin{itemize}
\item An interleaving $p$ is a sequence of memory accesses denoted
  $p$($sch$, $in$) = $\{A_1,...,A_n\}$ where $sch$ represents the
  order of these accesses and $in$ represents the data input.
\item Each $A_i$, where $i\in [0,n]$, denotes a memory access. 
\item $pcon_i(k)$ is the path condition under which $A_i$ is reached.
\end{itemize}

Thus, when $pcon_i(k)$ is true, meaning $A_i$ is reachable, we check
if $A_i$ can lead to a cache hit:
\begin{itemize}
\item $\tau_{i}$($k$) denotes the condition under which $A_i$ triggers
  a cache hit.
\item $addr_i$ denotes the memory address accessed by $A_i$.
\item $tag(addr)$ is a function that returns the unique $tag$ of $addr$.
\item $line(addr)$ is a function that returns the cache line of $addr$.
\end{itemize}

Thus, we define the \emph{cache-hit condition} as follows:
\begin{multline}
\label{eqn:tau}
\tau_i (k) \equiv \bigvee_{0\leq j < {i}} 
~~\Big (~ tag(addr_j)=tag(addr_i) ~\wedge \\
\forall l\in \left[j\mathit{+}1,i\mathit{-}1\right]\big | line(addr_l)
\ne line(addr_i)\Big )  
\end{multline}
For each memory access $A_i$, $\SymSC$ traverses the preceding memory
accesses in the interleaving $p$ to see if any such $A_j$ may result
in $A_i$ being a cache hit.  This is done by comparing the tag of
$addr_i$ to that of $addr_j$---a hit is possible only when two
tags are the same.  Furthermore, any other memory access ($A_l$)
between $A_i$ and $A_j$ must not evict the cache line occupied by
$A_j$ (and hence $A_i$).  This means, for all $j<l<i$, we have
$line(addr_l) \ne line(addr_i)$.

If $A_i$ always causes a cache hit, or a miss, it cannot leak
sensitive information because it implies $\forall k_1,k_2 ~.~
\tau_i(k_1) = \tau_i(k_2)$.  In contrast, if $\tau_i(k)$ evaluates to
\emph{true} for some value of $k$ but to \emph{false} for a different
value of $k$, then it is a leak.

\subsection{Leakage Detection}

After constructing $\tau_i(k)$, which is the \emph{cache-hit
  condition} for a potentially adversarial memory access $A_i$, we
instantiate the symbolic expression twice, first with a fresh variable
$k_1$ and then with another fresh variable $k_2$.  We use an
off-the-shelf SMT solver to search for values of $k_1$ and $k_2$ that
can lead to divergent behaviors.

\paragraph{Precise Solution}
The precise formulation is as follows:
\begin{equation}
\label{eqn:precise}
\exists k_1,k_2~.~(k_1 \neq k_2) \wedge \tau_i(k_1) \neq \tau_i(k_2)  
\end{equation}
We need to conduct this check at every memory access $A_i$, where
$i\in [0,n]$, along the symbolic execution path $p$.  If the above
formula is satisfiable, the SMT solver will return values
$\overline{k_1}$ and $\overline{k_2}$ of variables $k_1$ and $k_2$,
respectively.

\paragraph{Two-Step Approximation}

Since computing both values at the same time is expensive, in
practice, we can take two steps:
\begin{itemize}
\item 
First, solve subformula $\exists k_1 ~.~ \tau_i(k_1)$ to compute a
concrete value for $k_1$, denoted $\overline{k_1}$.
\item 
Second, solve subformula $\exists k_2~.~(\overline{k_1} \neq k_2)
\wedge \tau_i(\overline{k_1}) \neq \tau_i(k_2)$ to compute a 
concrete value $\overline{k_2}$ for $k_2$.
\end{itemize}
Since the formula solved in each step is (almost twice) smaller, the
solving time can be reduced significantly.  Furthermore, a valid solution
($\overline{k_1}$ and $\overline{k_2}$) is guaranteed to be a valid
solution for the original formula as well.
However, in general, the two-step approach is an under-approximation:
when it fails to find any solution, it is not a proof that no such
solution exists.

To make the two-step approach precise, one would have to apply it
repeatedly, each time with a different $\overline{k_1}$ computed in
the first step, until all solutions of $\overline{k_1}$ is covered.
Nevertheless, we shall show through experiments that, in practice,
applying it once is often accurate enough to detect the actual leak.

\subsection{The Running Example}
\label{sec:motivating_revisiting}

We revisit the example in Figure~\ref{fig:motiv-conc} to show how our
approach detects the leak.  Recall that \SymSC{} would generate the
six interleavings shown in Table~\ref{tbl:interleavings}.  For each
interleaving, Table~\ref{tbl:individual_constraint} shows the line
number (\#line) of every access $A_i$, path condition $pcon_i$, memory
address $addr_i$, and the cache-hit constraint $\tau_i$.

\begin{table}
\vspace{1ex}
  \caption{Cache-related information of interleaving $p$.}
  \label{tbl:individual_constraint}
  \centering
  \resizebox{.99\linewidth}{!}{
    \begin{tabular}{|c|c|c|c|c|c|}
      \hline
 	  {~~\# line~~} & ~~~$i$~~~ & $pcon_i$      & {$addr_i$} & {$\tau_i$}                  & {~~cache~~}  \\\hline\hline

	  {\tt 6}  &  0  & $k\leq 127$ & q[255-k]   & $\mathit{false}$           & miss  \\\hline
 	  {\tt 9}  &  1  & $k\leq 127$ & p[k]       & $~tag(p[k])=tag(q[255-k])$ & miss  \\\hline
	  {\multirow{3}{*}{\tt 13}} 
                   &    & \multirow{3}{*}{$k\leq 127$}
                                       & \multirow{3}{*}{tmp}
                                                    & {$tag(tmp)=tag(p[k])\vee$} & {\multirow{3}{*}{}}\\
 		   &  2  &             &            & $\big(tag(tmp)=tag(q[255-k])$ & miss\\
 		   &     &             &            &$\wedge line(tmp)\neq line(p[k]) \big)$ & \\\hline
 	  {\multirow{3}{*}{\tt 11}} 
                   &     & \multirow{3}{*}{$k\leq 127$}
                                       & \multirow{3}{*}{p[k]}
                                                    & {$tag(p[k])=tag(tmp)\vee$} & {\multirow{3}{*}{}}miss\\
 		   &  3   &             &            & $\big(tag(p[k])=tag(p[k])$ &   \textit{or} \\
 		   &     &             &            & $\wedge~line(p[k])\neq line(tmp) \big)$ & hit \\
       \hline
    \end{tabular}
  }
\vspace{-2ex}
\end{table}

Inside the interleaving \texttt{6-9-13-11}, for instance, upon
reaching the $\texttt{load}$ of \texttt{q[255-k]} at line 6, the path
condition would be ($\texttt{k} \leq \texttt{127}$).  Since it is the
first memory access, $\tau_0$ must be $\mathit{false}$ (cache miss).
We will record this memory address for further analysis.

Next is the $\texttt{load}$ of $\texttt{p[k]}$ at line 9.  $\SymSC$
builds $\tau_1$ and checks its satisfiability.  Since $\texttt{p[k]}$
and the preceding $\texttt{q[255-k]}$ correspond to different memory
addresses, the $tag$ comparison in $\tau_1$ returns $\mathit{false}$,
indicating a cache miss.  The $\texttt{load}$ at line 13 accesses
$\texttt{tmp}$.  Since $\texttt{tmp}$ is different from any of the
elements in arrays $\texttt{p}$ and $\texttt{q}$, the $tag$
comparisons in $\tau_2$ return \textit{false}, making $A_2$ a cache
miss.

Similarly, $\tau_3$ for the $\texttt{store}$ at line 11 is shown in
the last row of Table~\ref{tbl:individual_constraint}.  It is worth
mentioning that $\tau_3$ only compares $\texttt{p[k]}$ ($addr_3$) with
$\texttt{tmp}$ ($addr_2$) and $\texttt{p[k]}$ ($addr_1$) but not
$\texttt{q[255-k]}$ ($addr_0$) because $\SymSC$ finds that, if the
access to $\texttt{tmp}$ does not evict the cache line used by its
preceding access to $\texttt{p[k]}$ ($addr_1$), the last
$\texttt{store}$ to $\texttt{p[k]}$ ($addr_3$) must be a cache hit;
$\SymSC$ stops here to avoid further (and unnecessary) analysis.

Differing from $\tau_0$, $\tau_1$ and $\tau_2$, the constraint
$\tau_3$ depends on $k$ due to the constraint $line(p[k])\neq
line(tmp)$.  Specifically, $\tau_3(k)$ is $\mathit{true}$ when
($k!=1\wedge k\leq 127$) and is \emph{false} when ($k=1$).

In \SymSC{}, two symbolic variables $k_1,k_2$ will be used to
substitute $k$ in the symbolic expression of $\tau_3(k)$, to form
$\tau_3(k_1)$ and $\tau_3(k_2)$.  Solving the satisfiability problem
described by $\tau_3(k_1) ~XOR~ \tau_3(k_2)$ would produce the
assignment $\{ k_1\texttt{=0}$ and $k_2\texttt{=1} \}$, which makes
$\tau_3(0)$ evaluate to \emph{true} and $\tau_3(1)$ evaluate to
\emph{false}.

\section{Optimizations}
\label{sec:optimization}

Symbolic execution, when applied directly to cipher programs, may have
a high computational overhead because of the heavy use of arithmetic
computations and look-up tables in these programs.  In this section,
we present techniques for reducing the overhead.

Toward this end, we have two insights.
First, when conducting cache analysis, we are not concerned with the
actual numerical computations inside the cipher unless they affect the
addresses of memory accesses that may depend on sensitive data,
e.g., indices of lookup tables such as S-Boxes.
Second, for the purpose of detecting leaks, as opposed to proving
their absence, we are free to under-approximate as long as it does not
diminish the leak-detection capability of our analysis.

\subsection{Domain-specific Reduction}
\label{sec:domain_study}

By studying real-world cipher programs, we have found the
computational overhead is often associated with symbolic indices of
lookup tables such as the one shown in Figure~\ref{fig:sbox}.

\begin{figure}
\vspace{1ex}
\begin{lstlisting}
uint8_t SBOX1[64]={0x6f,0x3c,0x77,0xb7,0x2f,0x7b,0x5f,0xc6, ...};
uint8_t SBOX2[64]={0x3d,0x4c,0x5f,0xb6,0xd1,0xff,0x3e,0xed, ...};
void encrypt(uint8_t *block){
	for (uint8_t i = 0; i < 64; i++){ 
		block[i] |= SBOX1[block[i]];
		block[i] ^= SBOX2[block[i]];
	}
}	
\end{lstlisting}
\vspace{-1ex}
\caption{Example code for accessing S-Box lookup tables.}
\label{fig:sbox}
\vspace{-2ex}
\end{figure}

Here, \texttt{block} points to a 8-byte storage area whose content
depends on the cryptographic key; thus, the eight bytes are
initialized with symbolic values.  Accordingly, indices to the S-Box
tables -- \texttt{block[i]} at line 4 -- are symbolic.  
However, not all memory accesses should be treated as symbolic.  For
example, the address of \texttt{block[i]} itself, and the address of
local variables such as \texttt{i} should be treated as concrete
values to reduce the cost of symbolic execution.  Therefore, we
conduct a static analysis of the interleaved execution trace $p$ to
identify the sequence of memory accesses that need to be kept
symbolic while avoiding the symbolic expressions of other unnecessary
memory addresses.

Also, a program may have multiple S-Box arrays, like \texttt{SBOX1}
and \texttt{SBOX2} in Figure~\ref{fig:sbox}.  Two successive accesses
to \texttt{SBOX1} and \texttt{SBOX2} (at lines 5 and 6) cannot form a
cache hit no matter what the lookup indices are.  Therefore, we do not
need to invoke the SMT solver to check the equivalence of these
symbolic addresses.  This can significantly cut down the
constraint-solving time.

\subsection{Layout-directed Reduction}
\label{sec:approx}

Another reduction is guided by the memory layout.  In LLVM, memory
layout may be extracted from the compiler back-end after the code
generation step.  Recall that when analyzing a pair of potentially
adversarial addresses, we need to compare them with all other
addresses accessed between them to build the cache behavior
constraint.
More specifically, to check if \texttt{$A_2$} is a cache hit because
of $A_1$ along the execution $A_1-B_1-,...,-B_n-A_2$, we need to check
if any $B_i$ ($1\leq i\leq n$) could evict the cache line used by
$A_1$.  Due to the large value of $n$ and often complex symbolic
expression of $B_i$, the constraint-solving time could be large.

Our approach in this case is to directly compare $A_1$ and $A_2$ while
postponing the comparisons to $B_i$.  This is based on the observation
that, in practice, the cache line of $A_1$ can possibly be evicted by
$B_i$ only if the differences between their addresses is the multiple
of the cache size (e.g., 64KB), which may not be possible in compact
cipher programs.  For example, in a 64KB direct-mapped cache, for
$B_1$ to evict the 64-byte cache line of $A_1$, their address
difference has to be $2^{16}=64$KB.  In a 4-way set-associative cache,
their address difference has to be $2^{14}=16$KB.
Furthermore, in the event that $A_2$ has a cache hit due to $A_1$, we
can add back the initially-omitted comparisons to $B_1,\ldots,B_n$ to
undo the approximation.

\ignore{

\subsection{Early Termination}
\label{sec:early}

In Figure~\ref{fig:specific_interleaving}, the secondary thread $T_2$
preempts $T_1$ before $T_1$'s last store event, causing side-channel
leak when the control comes back to $T_1$ and executes the store.
Although in this particular example, $T_1$ ends right after executing
the store instruction, in practice, $T_1$ may continue to execute many
other events. However, since the adversarial access from $T_2$ has
been executed, continuing with the subsequent events in $T_1$ no
longer benefits our analysis.

Therefore, we make \SymSC{} perform an early termination (i.e.,
backtracking) upon witnessing a timing leak in the current
interleaving.  For instance, \texttt{6-13-9-11} and \texttt{8-13-9-11}
in Table~\ref{tbl:interleavings} can be terminated immediately before
executing the third event at Line~9 thus avoiding the additional
overhead.  It is worth noting that early termination only skips the
unnecessary explorations w.r.t to detecting timing leaks.  It will not
miss leaks because other possible leakage points (skipped by early
termination) will eventually be detected in other interleavings as
their first leakage points.

}

\section{Experiments}
\label{sec:evaluation}

We have implemented \SymSC{} using the LLVM compiler~\cite{LattnerA04}
and \emph{Cloud9}~\cite{BucurUZC11}, which is a symbolic execution
engine for multithreaded programs built upon KLEE~\cite{CadarDE08}.
We enhanced \emph{Cloud9} in three aspects.  
First, we extended its support for multi-threading by allowing context
switches prior to accessing global memory; the original \emph{Cloud9}
only allows context switches prior to executing a synchronization
primitive (e.g., lock/unlock).  
Second, we made \emph{Cloud9} fork new states to flip the execution
order of two simultaneously enabled events when they may be mapped to
the same cache line; the original \emph{Cloud9} does not care about
cache lines.  
Third, we made \emph{Cloud9} record the address of each memory access
along the execution, so it can incrementally build the cache-hit
constraint. 
Based on these enhancements, we implemented our cache timing leak
detector and optimized it for efficient constraint solving.

After compiling the C code of a program to LLVM bit-code, our \SymSC{}
tool executes it symbolically to generate interleavings according to
Algorithm~\ref{alg:symsc}.  The cache constraint at each memory access
is expressed in standard KQuery expressions defined in
KLEE~\cite{CadarDE08}.  By solving these constraints, we can obtain a
concrete execution that showcases the leak, including a thread
schedule, two input values $\overline{k_1},\overline{k_2}$ and the
adversarial memory address.

\subsection{Benchmarks}
\label{sec:benchs}

We evaluated \SymSC{} on a diverse set of open-source cipher programs.
Specifically, the first group has five programs from a lightweight
cryptographic system named \emph{FELICS}~\cite{DinuBGKCP2015}, which
was designed for resource-constrained devices.  The second group has
four programs from \emph{Chronos}~\cite{Chronos}, a real-time Linux
kernel.  The third group has four programs from the GNU 
library \emph{Libgcrypt}~\cite{Libgcrypt}, while the remaining
programs are from the \emph{LibTomCrypt}~\cite{LibTomCrypt}, the
\emph{OpenSSL}~\cite{OpenSSL}, and a recent
publication~\cite{Chattopadhyay17}.  They include multiple versions of
several well-known algorithms such as
\texttt{AES}~\cite{OpenSSL,Chronos} and
\texttt{DES}~\cite{Libgcrypt,Chronos}, which are useful in evaluating
the impact of cipher implementations on the performance of \SymSC{}.

\begin{table}
\vspace{1ex}
\caption{\fontsize{8}{11}\selectfont Benchmark statistics: lines of C code (LOC) and LLVM code (LL), sensitive key-size (KS), and the memory accesses (MA).}
\label{tbl:benchs}
\centering
\resizebox{\linewidth}{!}{
\begin{tabular}{|l|r|r|r|r||l|r|r|r|r|} \hline
\textbf{Name} &\textbf{LOC}&\textbf{LL} &\textbf{KS} &\textbf{MA} &\textbf{Name}&\textbf{LOC}&\textbf{LL}&\textbf{KS}&\textbf{MA}\\ 
\hline
\hline
\texttt{AES}\cite{OpenSSL} & 1,429 & 4,384& 24 &771&
\texttt{FCrypt}\cite{Chronos}& 437  & 1,623&12&428\\ 

\hline
\texttt{AES}\cite{Chronos} & 1,368 &4,144 & 24 &788& 
\texttt{KV\_name}\cite{Chattopadhyay17} &1,350&1,402&4&19 \\

\hline
\texttt{Camellia}\cite{LibTomCrypt} & 776 & 5,319& 16&1,301&
\texttt{LBlock}\cite{DinuBGKCP2015}& 930&4,010&10&1,618\\

\hline
\texttt{CAST5}\cite{LibTomCrypt} & 735 & 2,790& 16&909&
\texttt{Misty1}\cite{Botan}&391&1,199&16&270\\

\hline
\texttt{CAST5}\cite{Chronos} & 883 & 4,190& 16&1,180&
\texttt{Piccolo}\cite{DinuBGKCP2015}&301&1,034&12&350\\

\hline
\texttt{Chaskey}\cite{DinuBGKCP2015}& 248  & 638&16&242& 
\texttt{PRESENT}\cite{DinuBGKCP2015}&194&272&10&94\\

\hline
\texttt{DES}\cite{Libgcrypt}& 596  & 2,166&8&963 &
\texttt{rfc2268}~\cite{Libgcrypt}&388&870&16&149\\  

\hline
\texttt{DES}\cite{Chronos}& 1,010  & 3,926&8&1,029&
\texttt{Seed}\cite{Libgcrypt}&607&3,535&16&979\\ 

\hline
\texttt{Kasumi}\cite{Botan} & 350&1224&16&259&
\texttt{TWINE}\cite{DinuBGKCP2015}&256&562&10&229\\ 

\hline
\texttt{Khazad}\cite{Chronos} & 838&463&16&123&
\texttt{Twofish}\cite{Libgcrypt} &1,048&4,510&16&1,180\\ 
\hline
\end{tabular}
}
\vspace{-2ex}
\end{table}

Table~\ref{tbl:benchs} shows the statistics of these benchmark
programs.  The \textbf{LOC} and \textbf{LL} columns denote the lines
of C code and the corresponding LLVM bit-code.  The \textbf{KS} column
shows the size of the sensitive input in bytes.  The maximum number of
memory accesses on program paths of each benchmark is shown in the
\textbf{MA} column, which indicates the computational cost of the
program.

Each program in the benchmark suite has from 194 to 1,429 lines of C
code.  In total, there are 14,455 lines of C code, which compile to
49,048 lines of LLVM bit-code.  These numbers are considered
substantial because ciphers are typically compact programs with highly
computation-intensive operations, e.g., due to their use of loops and
lookup-table based transformations.  For example, the program named
\texttt{PRESENT} has only 194 lines of C code but 8,233 memory
accesses at run time.


We analyzed these benchmark programs using two types of caches:
direct-mapped cache and four-way set-associative cache.  The cache
size is 64KB with each cache line consisting of 64 bytes; thus, there
are 64KB/64B = 1024 cache lines, which are typical in mainstream
computers today.

Our experiments were designed to answer two questions:
\begin{compactitem}
\item 
Can \SymSC{} detect cache-timing leaks exposed by concurrently running
a program with other threads?
\item 
Are the optimizations in Section~\ref{sec:optimization} effective in
reducing the cost of symbolic execution and constraint solving?
\end{compactitem}
We conducted all experiments with Ubuntu 12.04 Linux running on a
computer with a 3.40GHz CPU and 8GB RAM.  For all evaluations we set
the timeout threshold to 1,600 minutes.

\subsection{Results Obtained with Fixed Addresses}

\begin{table}
\vspace{1ex}
\caption{Results of leak detection with \emph{fixed} addresses: Is the program leaky w.r.t. the \emph{given} thread?}
\label{tbl:Fixed_addr}
\centering
\resizebox{\linewidth}{!}{
\begin{tabular}{|l|r|c|r|r|c|r|}
\cline{1-7}
\multirow{3}{*}{\textbf{Name}} 		&  
\multicolumn{3}{c|}{\textbf{Precise}}&		
\multicolumn{3}{c|}{\textbf{Two-Step}}\\
\cline{2-7}
      &\multirow{2}{*}{~~\#.Inter}   
                    &\multirow{2}{*}{~~\#.Test}  
                                &\multirow{2}{*}{~~Time (m)} 
                                            &\multirow{2}{*}{~~\#.Inter} 
                                                        &~~\#.Test        &\multirow{2}{*}{~~Time (m)} \\ 
\cline{6-6}
      &             &           &           &           & step1~/~step2   &           \\ 
\hline\hline
AES\cite{OpenSSL}       	&57  &55   &430.2  &57  &55~/~55  &140.3\\  
AES\cite{Chronos}    		&1   &0    &288.9  &1   &1~/~0  &41.4\\ 
Camellia\cite{LibTomCrypt}  	&1   &0    &0.1    &1   &1~/~0  &0.1\\ 
CAST5\cite{LibTomCrypt}         &1   &0    &0.1    &1   &1~/~0 &0.1\\ 
CAST5\cite{Chronos}       	&1   &0    &0.1    &1   &1~/~0     &0.1\\   
Chaskey\cite{DinuBGKCP2015}     &1   &0    &0.1    &1   &1~/~0    &0.1\\   
DES\cite{Libgcrypt}   		&16  &15   &7.8    &16  &16~/~15    &3.5\\
DES\cite{Chronos}    		&1   &0    &0.1    &1   &1~/~0    &0.1\\ 
FCrypt\cite{Chronos}       	&16  &15   &4.1    &16  &15~/~15   &8.1\\
Kasumi\cite{Botan} 		&1   &0    &0.1    &1   &1~/~0    &0.2\\ 
Khazad\cite{Chronos}    	&25  &23   &206.5  &25  &23~/~23    &83.0\\ 
KV\_Name\cite{Chattopadhyay17}~~~~~  
                                &1406&0    &0.5    &1406&1406~/~0    &0.4\\ 
LBlock\cite{DinuBGKCP2015}      &1   &0    &0.1    &1   &1~/~0    &0.1\\ 
Misty1\cite{Botan} 		&1   &0    &0.1    &1   &1~/~0    &0.1\\ 
Piccolo\cite{DinuBGKCP2015}     &1   &0    &0.1    &1   &1~/~0    &0.1\\ 
PRESENT\cite{DinuBGKCP2015}     &1   &0    &0.1    &1   &1~/~0    &0.1\\ 
rfc2268\cite{Libgcrypt}     	&1   &0    &0.1    &1   &1~/~0    &0.1\\ 
Seed\cite{Libgcrypt}     	&1   &0    &0.1    &1   &1~/~0    &0.1\\ 
TWINE\cite{DinuBGKCP2015}       &1   &0    &0.1    &1   &1~/~0    &0.1\\ 
Twofish\cite{Libgcrypt}         &1   &0    &0.1    &1   &1~/~0    &0.2\\ 
\hline
\end{tabular}
}
\vspace{-1ex}
\end{table}

Table~\ref{tbl:Fixed_addr} shows our results obtained using fixed
addresses in the cache layout (Case 1 in
Section~\ref{sec:app-scenarios}).  The first column shows the
benchmark name.  The next three columns show the result of computing
the precise solution for our cache analysis problem.  The last three
columns show the result of running the simplified, two-step version,
where the solution for $\exists k_1,k_2 ~.~ \tau(k_1) \neq \tau(k_2)$
is computed in two steps, by first computing a value of $k_1$ and then
computing a value of $k_2$.  In each method, we show the number of
interleavings explored (\#.Inter), the number of leaky memory accesses
detected (\#.Test), and the execution time in minutes (m).  For the
two-step approach, we also show the number of leakage points detected
after the first step and after the second step.

Among these twenty programs, we detected leakage points in four:
\texttt{ASE} from \emph{OpenSSL}~\cite{OpenSSL}, \texttt{DES} from
\emph{Libgcrypt}~\cite{Libgcrypt}, \texttt{FCrypt} from
\emph{Chronos}~\cite{Chronos}, and \texttt{Khazad} from
\emph{Chronos}~\cite{Chronos}. We manually inspected these four
programs in a way similar to what is described in
Section~\ref{sec:hpn-ssh}, and confirmed that all these leakage points
are realistic.  Furthermore, our two-step approach returned exactly
the same results as the precise analysis for all benchmark programs,
but in significantly less time.

We also conducted our experiments using \emph{4-way set-associative}
cache instead of \emph{direct-mapped} cache. The results of these
experiments are similar to the ones reported in Table
\ref{tbl:Fixed_addr}.
Therefore, we omit them for brevity.

Nevertheless, the similarity is expected.  For example, a 1024-byte
S-Box would be mapped to 16 consecutive cache lines in
\emph{directed-mapped} cache as well as \emph{4-way set-associative}
cache, provided that the cache size is 64KB and the line size is
64-byte.  The only minor difference is that, in the \emph{4-way
  set-associative} cache, we need four adversarial memory accesses
from thread $T_2$ to fully evict a cache set.  But if we have already
detected the first adversarial address (say \textit{addr}), the
remaining three could simply be \textit{addr+cache\_size},
\textit{addr+2*cache\_size}, and \textit{addr+3*cache\_size}.  Thus,
there is no significant difference from analyzing \emph{direct-mapped}
cache.

\subsection{Results Obtained with Symbolic Addresses}
\label{sec:results}

The results shown in Table~\ref{tbl:Fixed_addr} are useful, but also
somewhat \emph{conservative}.  A more aggressive analysis is to assume
the adversarial thread $T_2$ may access memory regions whose cache
layout is symbolic (refer to Case 2 in
Section~\ref{sec:app-scenarios}).

\begin{table}
\vspace{1ex}
\caption{Results of leak detection with \emph{symbolic} addresses: Is
  the given program leaky w.r.t.  \emph{any} adversarial thread?}
\label{tbl:direct_updated}
\centering
\resizebox{\linewidth}{!}{
\begin{tabular}{|l|r|r|c|r||r|c|r|}
\hline
\multirow{3}{*}{\textbf{Name}} &
\multirow{3}{*}{\textbf{\#.Acc}} &
\multicolumn{3}{c|}{\textbf{Precise}} &
\multicolumn{3}{c|}{\textbf{Two-Step}} \\
\cline{3-8}
             &        &\multirow{2}{*}{\#.Inter}   
                                 &\multirow{2}{*}{~~\#.Test}  
                                            &\multirow{2}{*}{Time(m)} 
                                                       &\multirow{2}{*}{\#.Inter} 
                                                                  & ~~\#.Test        &\multirow{2}{*}{Time(m)} \\ 

\cline{7-7}
             &       &           &          &          &          & step1~/~step2    &         \\
\hline\hline
AES~\cite{OpenSSL}             &1,026  &224   &220  &1016.4  &224  &220~/~220  &237.5    \\
AES\cite{Chronos}              &2,568  &141   &139  &>1600  &256  &302~/~254 &548.3    \\
Camellia\cite{LibTomCrypt}     &2,590  &176   &172  &830.8  &176  &172~/~172  &303.5    \\
CAST5\cite{LibTomCrypt}        &1,815  &167   &164  &>1600  &384  &381~/~381  &1337.4   \\
CAST5\cite{Chronos}            &1,392  &183   &180  &>1600  &384  &381~/~381  &1392.5   \\
Chaskey\cite{DinuBGKCP2015}    &1,380  &1     &0    &0.1    &1    &1~/~0    &0.1      \\
DES\cite{Libgcrypt}            &2,135  &144   &127  &38.6    &144  &164~/~127 &27.2     \\
DES\cite{Chronos}              &2,539  &119   &114  &>1600  &194  &187~/~183  &1191.5   \\
FCrypt\cite{Chronos}           &428    &64    &60   &15.1    &64   &60~/~60   &20.1     \\
Kasumi\cite{Botan}             & 1,785 &83    &82   &>1600  &96   &94~/~94   &151.9    \\
Khazad\cite{Chronos}           &684    &114   &103  &>1600  &248  &254~/~240 &165.3    \\
KV\_Name\cite{Chattopadhyay17} &140    &1406  &0    &0.5    &1406 &1406~/~0 &0.5      \\
LBlock\cite{DinuBGKCP2015}     &4,068  &1     &0    &0.1    &1    &1~/~0    &0.1      \\
Misty1\cite{Botan}             &2,966  &76    &75   &>1600  &96   &94~/~94   &265.1    \\
Piccolo\cite{DinuBGKCP2015}    &5,103  &1     &0    &0.1    &1    &1~/~0    &0.1      \\
PRESENT\cite{DinuBGKCP2015}    &8,233  &1     &0    &0.2    &1    &1~/~0    &0.2      \\
rfc2268\cite{Libgcrypt}        &3,190  &113   &112  &303.4  &113  &112~/~112  &42.9     \\
Seed\cite{Libgcrypt}           &1,632  &201   &197  &>1600  &320  &316~/~316  &1505.1   \\
TWINE\cite{DinuBGKCP2015}      &10,492 &1     &0    &0.1    &1    &1~/~0    &0.1      \\
Twofish\cite{Libgcrypt}        &12,400 &2514  &84   &>1600  &900 &84,063~/~76&>1600   \\
\hline

\end{tabular}
}
\vspace{-1ex}
\end{table}

Table~\ref{tbl:direct_updated} shows the experimental results obtained
using direct-mapped cache and symbolic addresses in thread $T_2$
(Case 2 in Section~\ref{sec:app-scenarios}).
The first two columns show the benchmark name and the maximum number
of memory addresses accessed by an interleaving at run time.
The \textit{Precise} column shows the result of computing the precise
solution for our cache analysis problem.
The \textit{Two-Step} column shows the result of running the
simplified version. 
In both cases, we report the total number of interleavings explored by
symbolic execution (\#.Inter), the number of leaky memory accesses
detected (\#.Test), and the total execution time in minutes (m).
For \textit{Two-Step}, the number of leaky accesses is further divided
into two subcolumns: the leaky accesses detected after the first step
and the leaky accesses detected after the second step.

The results show that, for most of the benchmark programs, the
overhead of precisely solving our cache analysis is too high: on nine
of the twenty programs, it could not complete within the time limit.
In contrast, our two-step analysis was able to complete nineteen out of
the twenty programs.
In terms of accuracy, our two-step approach is almost as good as
precise analysis: in all completed programs, they detected the same
number of leakage points, which indicate a possible combination of
adversarial threads and memory layout that \emph{can} trigger timing
leaks.

Our results also show that, for the same type of cryptographic
algorithms (such as AES), different implementations may lead to
drastically different overhead.
For example, we detected 34 more leakage points in the \texttt{AES}
implementation of \emph{Chronos}~\cite{Chronos} than that of
\emph{OpenSSL}~\cite{OpenSSL}. However, the \texttt{AES} of
\emph{Chronos} took almost twice as long for our tool to analyze.
For \texttt{DES} the implementations from
\emph{Libgcrypt}~\cite{Libgcrypt} and \emph{Chronos}~\cite{Chronos},
we detected a slightly different number of leakage points, but the
time taken is significantly different (27.1 minutes versus 1191.5
minutes).
In contrast, for the two versions of \texttt{CAST5}, we detected the
same number of leakage points in roughly the same amount of time.

\ignore{
For the benchmark where \textit{Two-Step} took a long time, we
found the reason is because \SymSC{} keeps forking new symbolic states
at a branch point inside a loop, since the 10-byte symbolic input
makes both $\mathit{true}$ and $\mathit{false}$ conditions feasible,
which makes \emph{Cloud9} keep branching out.
}

For the benchmark where \textit{Two-Step} took a long time, we found
it is due to the increasing size of symbolic constraints which consist
of the addresses in S-Box accesses.  Typically the later a S-Box
access in a loop, the larger its symbolic address expression would be.
In \texttt{Twofish}, \SymSC{} timed out because it encountered a large
number of "may-be-related" event pairs (i.e., accessing the same S-Box
but not the same cache line), which made SMT solving difficult.

\subsection{Discussion}

Based on the results, we answer the two research questions as follows.
First, \SymSC{} is able to identify cache timing leaks in concurrent
programs automatically.  Specifically, using \emph{symbolic} addresses
in the adversarial thread allows us to demonstrate the possibility of
triggering leaks in a concurrent system, whereas using \emph{fixed}
addresses in the analysis allows us to show that such leaks are more
practical.
Second, \SymSC{}'s performance optimization techniques are effective
in reducing the computational overhead, which is demonstrated on a
diverse set of real-world cipher programs.

\SymSC{} searches for sensitive inputs as well as an interleaving
schedule that, together, trigger divergent cache behaviors.  If an
individual program path has a constant cache behavior, e.g., all the
memory accesses refer to fixed memory addresses regardless of the
value of the sensitive input, then timing leaks are impossible. 
By checking for and leveraging such conditions, \SymSC{} can reduce
the computation cost even further.
For instance, with naive exploration, \SymSC{} would have generated
1,406 interleavings for the benchmark program named
\texttt{KV\_name}. However, with the above analysis, it does not have
to generate any interleaving.

In this example, \texttt{KV\_name}'s 4-byte symbolic input only
affects the branch conditions but does not taint any memory access
address. Thus, many paths are explored by symbolic execution. However,
no leak is detected on these paths.

Another example is \texttt{Chaskey}, which has a single program path,
together with 1,380 memory accesses on this path.  These memory
addresses are all independent of the 16-Byte symbolic input, which
means no leakage point can be found by \SymSC{}.

\ignore{
In this work, we are concerned with \emph{detecting leaks} only as
opposed to \emph{constructing attacks} or \emph{quantifying the
  difficulty in constructing attacks}.  Nevertheless, the latter may
be addressed by conducting a quantitative version of our analysis,
e.g., by verifying $\forall k_1,k_2~.~|\tau(k_1)-\tau(k_2)|\leq
\Delta$ instead of $\forall k_1,k_2~.~\tau(k_1) = \tau(k_2)$.
However, we leave this for future work.
}

\section{Related Work}
\label{sec:related}

Side-channel leaks have been exploited in a wide range of
systems~\cite{Kocher96,DhemKLMQW98,KocherJJ99,MangardOP07,GenkinST14,GandolfiMO01,Quisquater2001,KongASZ13,GrussLSOHC17,MulderES18}.
For timing side channels, in particular, many analysis and verification
techniques have been developed.
For example, Chen et al.~\cite{ChenFD17} proposed a technique named
Cartesian Hoare Logic~\cite{SousaD16} for proving that the timing
leaks of a program are bounded.
Antonopoulos et al.~\cite{AntonopoulosGHK17} proposed a similar method
for proving the absence of timing channels: it partitions the program
paths in a way that, if individual partitions are proved to be timing
attack resilient, the entire program is also timing attack resilient.
However, these methods only consider \emph{instruction-induced} timing
while ignoring the cache.

In the context of analyzing real-time systems, there is a large body
of work on cache analysis~\cite{MitraTT18a,LiMR03,LiSLMR09}, with the goal of estimating
the worst-case execution time (WCET).  Various techniques including
abstract interpretation~\cite{TouzeauMMR17}, symbolic
execution~\cite{BasuC17,Chattopadhyay17}, and
interpolation~\cite{ChuJM16} have been used to compute the upper bound
of execution time along all program paths.
Chattopadhyay et al.~\cite{ChattopadhyayBRZ17} also developed
\textit{CHALICE} to quantify information leaked through the cache side
channel, but the focus was on dependencies between sensitive data and
misses/hits on the CPU's data cache.  

Doychev et al.~\cite{DoychevFKMR13} developed \textit{CacheAudit}, a
tool relying on abstract interpretation based static analysis to
analyze cache timing leaks.
Wang et al.~\cite{WangWLZW17} developed \textit{CacheD}, an offline
trace analysis tool for detecting key-dependent program points in a
cipher program that may be vulnerable to side channel attacks.
Sung et al.~\cite{SungPW18} developed \textit{CANAL}, an LLVM
transformation that models cache timing behaviors for standard
verification tools.
However, these techniques handle sequential programs or traces only.

Pasareanu et al.~\cite{PasareanuPM16} developed a symbolic execution
tool for reasoning about the degree of leaked information, assuming
the attacker can take multiple measurements.  The test input that
causes the maximum amount of leakage is computed using Max-SMT
solving.
Bultan et al.~\cite{BultanYAA17,BangAPPB16,BrennanSB18} developed
techniques for quantifying information leaked by string operations.
Their method can handle both single and multiple
runs~\cite{BangAPPB16}: it applies probabilistic symbolic execution to
collect path constraints of a single run and then uses these
constraints to compute the leakage of multiple runs.
Phan et al.~\cite{PhanBPMB17} also developed a symbolic attack model
and formulated the problem of test generation to obtain the maximum
leakage as an optimization problem.

However, in all these existing methods, the program is assumed to be
sequential. In contrast, \SymSC{} focuses on concurrency-induced
leaks.
Although Barthe et al.~\cite{BartheKMO14} proposed an abstract
interpretation technique based on
\textit{CacheAudit}~\cite{DoychevFKMR13} to track the cache state of a
program with concurrent adversary, the adversary is a separate process
(that tries to probe and set the cache states), not a thread.
Furthermore, users have to provide data inputs and interleaving
schedules, whereas \SymSC{} generates them automatically.

Stefan et al.~\cite{StefanBYLTRM13} proposed an instruction-based
scheduling mechanism in information flow control systems running on a
single CPU, to avoid cache timing attacks introduced by classic
time-based schedulers.  Therefore, it is a system-level mitigation
technique.  In contrast, \SymSC{} focuses on
detecting whether a security-critical program may leak sensitive
information through the timing side channel due to interference from
other threads.

Our state-space reduction in \SymSC{} is related to partial order
reduction (POR)~\cite{FlanaganG05} in model checking, but with an
important difference.  In classic POR~\cite{AronisJLS18,ChengYW17,ZhangKW15,KusanoW14,KahlonWG09,WangYKG08}, one would typically select
representative interleavings from equivalence classes, which are
defined based on standard data-conflict and data-dependence relations.
However, in \SymSC{}, they must be broadened to also include
functionally-independent events that may access the same cache line.

%
So far, \SymSC{} focuses on cases where the adversarial thread flushes
a single cache line.  In the terminology of side-channel analysis,
this corresponds to \emph{first-order} attacks.  If, on the other
hand, the adversarial thread is capable of flushing multiple cache
lines, it may be more likely to trigger timing leaks.  Such cases
would be called \emph{high-order} attacks.  We leave the analysis of
\emph{high-order} attacks for future work.
%

Besides leak detection, there are side-channel leak mitigation
techniques that can generate countermeasures automatically, e.g.,
using compiler-like program
transformations~\cite{BayrakRBSI11,MossOPT12,AgostaBP12,WuGSW18} or
SMT solver based formal
verification~\cite{EldibWS14,EldibWTS14,ZhangGSW18,BloemGIKMW18} and
program synthesis~\cite{EldibW14,EldibWW16,WangS17,BlotYT17} techniques.
However, none of these emerging techniques was designed for, or
applicable to, cache timing side channels due to concurrency.

\section{Conclusions}
\label{sec:conclusion}

We have presented a symbolic execution method for detecting cache
timing leaks in a computation that runs concurrently with an
adversarial thread.  Our method systematically explores both thread
paths and their interleavings, and relies on an SMT solver to detect
divergent cache behaviors.  Our experiments show that real cipher
programs do have concurrency related cache timing leaks, and although
it remains unclear \emph{to what extent} such leaks are exploited in
practice, our method computes concrete data inputs and interleaving
schedules to demonstrate these leaks are realistic.  To the best of
our knowledge, this is the first symbolic execution method for
detecting cache timing side-channel leaks due to concurrency.

%

%

\clearpage\newpage
\bibliographystyle{plain}
\bibliography{tapsc}

\begin{thebibliography}{10}

\bibitem{Botan}
{\em {Botan}}.
\newblock \url{https://botan.randombit.net/}.

\bibitem{hpn-ssh}
{\em {High Performance SSH/SCP - HPN-SSH}}.
\newblock \url{https://www.psc.edu/hpn-ssh}.

\bibitem{Libgcrypt}
{\em {Libgcrypt}}.
\newblock \url{https://gnupg.org/software/libgcrypt/index.html}.

\bibitem{LibTomCrypt}
{\em {LibTomCrypt}}.
\newblock \url{http://www.libtom.net/LibTomCrypt/}.

\bibitem{openssh}
{\em {OpenSSH}}.
\newblock \url{http://www.openssh.com/}.

\bibitem{OpenSSL}
{\em {OpenSSL}}.
\newblock \url{https://github.com/openssl/openssl/tree/OpenSSL_0_9_7-stable}.

\bibitem{AgostaBP12}
Giovanni Agosta, Alessandro Barenghi, and Gerardo Pelosi.
\newblock A code morphing methodology to automate power analysis
  countermeasures.
\newblock In {\em ACM/IEEE Design Automation Conference}, pages 77--82, 2012.

\bibitem{AntonopoulosGHK17}
Timos Antonopoulos, Paul Gazzillo, Michael Hicks, Eric Koskinen, Tachio
  Terauchi, and Shiyi Wei.
\newblock Decomposition instead of self-composition for proving the absence of
  timing channels.
\newblock In {\em ACM SIGPLAN Conference on Programming Language Design and
  Implementation}, pages 362--375, 2017.

\bibitem{AronisJLS18}
Stavros Aronis, Bengt Jonsson, Magnus L{\aa}ng, and Konstantinos Sagonas.
\newblock Optimal dynamic partial order reduction with observers.
\newblock In {\em International Conference on Tools and Algorithms for
  Construction and Analysis of Systems}, pages 229--248, 2018.

\bibitem{BangAPPB16}
Lucas Bang, Abdulbaki Aydin, Quoc{-}Sang Phan, Corina~S. Pasareanu, and Tevfik
  Bultan.
\newblock String analysis for side channels with segmented oracles.
\newblock In {\em ACM SIGSOFT Symposium on Foundations of Software
  Engineering}, pages 193--204, 2016.

\bibitem{BartheKMO14}
Gilles Barthe, Boris K{\"{o}}pf, Laurent Mauborgne, and Mart{\'{\i}}n Ochoa.
\newblock Leakage resilience against concurrent cache attacks.
\newblock In {\em International Conference on Principles of Security and
  Trust}, pages 140--158, 2014.

\bibitem{BasuC17}
Tiyash Basu and Sudipta Chattopadhyay.
\newblock Testing cache side-channel leakage.
\newblock In {\em IEEE International Conference on Software Testing,
  Verification and Validation}, pages 51--60, 2017.

\bibitem{BayrakRBSI11}
Ali~Galip Bayrak, Francesco Regazzoni, Philip Brisk, Fran\c{c}ois-Xavier
  Standaert, and Paolo Ienne.
\newblock A first step towards automatic application of power analysis
  countermeasures.
\newblock In {\em ACM/IEEE Design Automation Conference}, pages 230--235, 2011.

\bibitem{BerganGC14}
Tom Bergan, Dan Grossman, and Luis Ceze.
\newblock Symbolic execution of multithreaded programs from arbitrary program
  contexts.
\newblock In {\em ACM SIGPLAN Conference on Object Oriented Programming,
  Systems, Languages, and Applications}, pages 491--506, 2014.

\bibitem{BloemGIKMW18}
Roderick Bloem, Hannes Gro{\ss}, Rinat Iusupov, Bettina K{\"{o}}nighofer,
  Stefan Mangard, and Johannes Winter.
\newblock Formal verification of masked hardware implementations in the
  presence of glitches.
\newblock In {\em Annual International Conference on the Theory and
  Applications of Cryptographic Techniques (EUROCRYPT)}, pages 321--353, 2018.

\bibitem{BlotYT17}
Arthur Blot, Masaki Yamamoto, and Tachio Terauchi.
\newblock Compositional synthesis of leakage resilient programs.
\newblock In {\em International Conference on Principles of Security and
  Trust}, pages 277--297, 2017.

\bibitem{BrennanSB18}
Tegan Brennan, Seemanta Saha, and Tevfik Bultan.
\newblock Symbolic path cost analysis for side-channel detection.
\newblock In {\em International Conference on Software Engineering}, pages
  424--425, 2018.

\bibitem{BucurUZC11}
Stefan Bucur, Vlad Ureche, Cristian Zamfir, and George Candea.
\newblock Parallel symbolic execution for automated real-world software
  testing.
\newblock In {\em European Conference on Computer Systems}, pages 183--198,
  2011.

\bibitem{BultanYAA17}
Tevfik Bultan, Fang Yu, Muath Alkhalaf, and Abdulbaki Aydin.
\newblock {\em String Analysis for Software Verification and Security}.
\newblock Springer, 2017.

\bibitem{CadarDE08}
Cristian Cadar, Daniel Dunbar, and Dawson~R. Engler.
\newblock {KLEE:} unassisted and automatic generation of high-coverage tests
  for complex systems programs.
\newblock In {\em USENIX Symposium on Operating Systems Design and
  Implementation}, pages 209--224, 2008.

\bibitem{Chattopadhyay17}
Sudipta Chattopadhyay.
\newblock Directed automated memory performance testing.
\newblock In {\em International Conference on Tools and Algorithms for
  Construction and Analysis of Systems}, pages 38--55, 2017.

\bibitem{ChattopadhyayBRZ17}
Sudipta Chattopadhyay, Moritz Beck, Ahmed Rezine, and Andreas Zeller.
\newblock Quantifying the information leak in cache attacks via symbolic
  execution.
\newblock In {\em ACM-IEEE International Conference on Formal Methods and
  Models for System Design}, pages 25--35, 2017.

\bibitem{ChenFD17}
Jia Chen, Yu~Feng, and Isil Dillig.
\newblock Precise detection of side-channel vulnerabilities using quantitative
  cartesian hoare logic.
\newblock In {\em ACM SIGSAC Conference on Computer and Communications
  Security}, pages 875--890, 2017.

\bibitem{ChengYW17}
Lin Cheng, Zijiang Yang, and Chao Wang.
\newblock Systematic reduction of {GUI} test sequences.
\newblock In {\em IEEE/ACM International Conference On Automated Software
  Engineering}, pages 849--860, 2017.

\bibitem{ChuJM16}
Duc{-}Hiep Chu, Joxan Jaffar, and Rasool Maghareh.
\newblock Precise cache timing analysis via symbolic execution.
\newblock In {\em IEEE Symposium on Real-Time and Embedded Technology and
  Applications}, pages 293--304, 2016.

\bibitem{CiorteaZBCC09}
Liviu Ciortea, Cristian Zamfir, Stefan Bucur, Vitaly Chipounov, and George
  Candea.
\newblock Cloud9: a software testing service.
\newblock {\em Operating Systems Review}, 43(4):5--10, 2009.

\bibitem{Chronos}
Matthew Dellinger, Piyush Garyali, and Binoy Ravindran.
\newblock Chronos linux: a best-effort real-time multiprocessor linux kernel.
\newblock In {\em ACM/IEEE Design Automation Conference}, pages 474--479, 2011.

\bibitem{DhemKLMQW98}
Jean{-}Fran{\c{c}}ois Dhem, Fran{\c{c}}ois Koeune, Philippe{-}Alexandre Leroux,
  Patrick Mestr{\'{e}}, Jean{-}Jacques Quisquater, and Jean{-}Louis Willems.
\newblock A practical implementation of the timing attack.
\newblock In {\em International Conference on Smart Card Research and
  Applications}, pages 167--182, 1998.

\bibitem{DinuBGKCP2015}
Daniel Dinu, Yann~Le Corre, Dmitry Khovratovich, Léo Perrin, Johann
  Großschädl, and Alex Biryukov.
\newblock Triathlon of lightweight block ciphers for the internet of things.
\newblock Cryptology ePrint Archive, Report 2015/209, 2015.

\bibitem{DoychevFKMR13}
Goran Doychev, Dominik Feld, Boris K{\"{o}}pf, Laurent Mauborgne, and Jan
  Reineke.
\newblock Cacheaudit: {A} tool for the static analysis of cache side channels.
\newblock In {\em USENIX Security Symposium}, pages 431--446, 2013.

\bibitem{EldibW14}
Hassan Eldib and Chao Wang.
\newblock Synthesis of masking countermeasures against side channel attacks.
\newblock In {\em International Conference on Computer Aided Verification},
  pages 114--130, 2014.

\bibitem{EldibWS14}
Hassan Eldib, Chao Wang, and Patrick Schaumont.
\newblock {SMT}-based verification of software countermeasures against
  side-channel attacks.
\newblock In {\em International Conference on Tools and Algorithms for
  Construction and Analysis of Systems}, pages 62--77, 2014.

\bibitem{EldibWTS14}
Hassan Eldib, Chao Wang, Mostafa Taha, and Patrick Schaumont.
\newblock {QMS}: Evaluating the side-channel resistance of masked software from
  source code.
\newblock In {\em ACM/IEEE Design Automation Conference}, pages 209:1--6, 2014.

\bibitem{EldibWW16}
Hassan Eldib, Meng Wu, and Chao Wang.
\newblock Synthesis of fault-attack countermeasures for cryptographic circuits.
\newblock In {\em International Conference on Computer Aided Verification},
  pages 343--363, 2016.

\bibitem{FlanaganG05}
Cormac Flanagan and Patrice Godefroid.
\newblock Dynamic partial-order reduction for model checking software.
\newblock In {\em ACM SIGACT-SIGPLAN Symposium on Principles of Programming
  Languages}, pages 110--121, 2005.

\bibitem{GandolfiMO01}
Karine Gandolfi, Christophe Mourtel, and Francis Olivier.
\newblock Electromagnetic analysis: Concrete results.
\newblock In {\em International Conference on Cryptographic Hardware and
  Embedded Systems}, pages 251--261, 2001.

\bibitem{GenkinST14}
Daniel Genkin, Adi Shamir, and Eran Tromer.
\newblock {RSA} key extraction via low-bandwidth acoustic cryptanalysis.
\newblock In {\em Annual International Cryptology Conference (CRYPTO)}, pages
  444--461, 2014.

\bibitem{GrussLSOHC17}
Daniel Gruss, Julian Lettner, Felix Schuster, Olga Ohrimenko, Istv{\'{a}}n
  Haller, and Manuel Costa.
\newblock Strong and efficient cache side-channel protection using hardware
  transactional memory.
\newblock In {\em USENIX Security Symposium}, pages 217--233, 2017.

\bibitem{GuoKW16}
Shengjian Guo, Markus Kusano, and Chao Wang.
\newblock {Conc-iSE}: incremental symbolic execution of concurrent software.
\newblock In {\em IEEE/ACM International Conference On Automated Software
  Engineering}, pages 531--542, 2016.

\bibitem{GuoKWYG15}
Shengjian Guo, Markus Kusano, Chao Wang, Zijiang Yang, and Aarti Gupta.
\newblock Assertion guided symbolic execution of multithreaded programs.
\newblock In {\em ACM SIGSOFT Symposium on Foundations of Software
  Engineering}, pages 854--865, 2015.

\bibitem{GuoWW17}
Shengjian Guo, Meng Wu, and Chao Wang.
\newblock Symbolic execution of programmable logic controller code.
\newblock In {\em ACM SIGSOFT Symposium on Foundations of Software
  Engineering}, 2017.

\bibitem{KahlonWG09}
Vineet Kahlon, Chao Wang, and Aarti Gupta.
\newblock Monotonic partial order reduction: An optimal symbolic partial order
  reduction technique.
\newblock In {\em International Conference on Computer Aided Verification},
  pages 398--413, 2009.

\bibitem{Kocher96}
Paul~C. Kocher.
\newblock Timing attacks on implementations of diffie-hellman, rsa, dss, and
  other systems.
\newblock In {\em Annual International Cryptology Conference (CRYPTO)}, pages
  104--113, 1996.

\bibitem{KocherJJ99}
Paul~C. Kocher, Joshua Jaffe, and Benjamin Jun.
\newblock Differential power analysis.
\newblock In {\em Annual International Cryptology Conference (CRYPTO)}, pages
  388--397, 1999.

\bibitem{KongASZ13}
Jingfei Kong, Onur Acii{\c{c}}mez, Jean{-}Pierre Seifert, and Huiyang Zhou.
\newblock Architecting against software cache-based side-channel attacks.
\newblock {\em {IEEE} Trans. Computers}, 62(7):1276--1288, 2013.

\bibitem{KopfMO12}
Boris K{\"{o}}pf, Laurent Mauborgne, and Mart{\'{\i}}n Ochoa.
\newblock Automatic quantification of cache side-channels.
\newblock In {\em International Conference on Computer Aided Verification},
  pages 564--580, 2012.

\bibitem{KusanoW14}
Markus Kusano and Chao Wang.
\newblock Assertion guided abstraction: a cooperative optimization for dynamic
  partial order reduction.
\newblock In {\em IEEE/ACM International Conference On Automated Software
  Engineering}, pages 175--186, 2014.

\bibitem{LattnerA04}
Chris Lattner and Vikram~S. Adve.
\newblock {LLVM:} {A} compilation framework for lifelong program analysis {\&}
  transformation.
\newblock In {\em IEEE/ACM International Symposium on Code Generation and
  Optimization}, pages 75--88, 2004.

\bibitem{LiMR03}
Xianfeng Li, Tulika Mitra, and Abhik Roychoudhury.
\newblock Accurate timing analysis by modeling caches, speculation and their
  interaction.
\newblock In {\em ACM/IEEE Design Automation Conference}, pages 466--471, 2003.

\bibitem{LiSLMR09}
Yan Li, Vivy Suhendra, Yun Liang, Tulika Mitra, and Abhik Roychoudhury.
\newblock Timing analysis of concurrent programs running on shared cache
  multi-cores.
\newblock In {\em IEEE Real-Time Systems Symposium}, pages 57--67, 2009.

\bibitem{MangardOP07}
Stefan Mangard, Elisabeth Oswald, and Thomas Popp.
\newblock {\em Power analysis attacks - revealing the secrets of smart cards}.
\newblock 2007.

\bibitem{MitraTT18a}
Tulika Mitra, J{\"{u}}rgen Teich, and Lothar Thiele.
\newblock Time-critical systems design: {A} survey.
\newblock {\em {IEEE} Design {\&} Test}, 35(2):8--26, 2018.

\bibitem{MossOPT12}
Andrew Moss, Elisabeth Oswald, Dan Page, and Michael Tunstall.
\newblock Compiler assisted masking.
\newblock In {\em International Conference on Cryptographic Hardware and
  Embedded Systems}, pages 58--75, 2012.

\bibitem{MulderES18}
Elke~De Mulder, Thomas Eisenbarth, and Patrick Schaumont.
\newblock Identifying and eliminating side-channel leaks in programmable
  systems.
\newblock {\em {IEEE} Design {\&} Test}, 35(1):74--89, 2018.

\bibitem{PasareanuPM16}
Corina~S. Pasareanu, Quoc{-}Sang Phan, and Pasquale Malacaria.
\newblock Multi-run side-channel analysis using symbolic execution and max-smt.
\newblock In {\em IEEE Computer Security Foundations Symposium}, pages
  387--400, 2016.

\bibitem{PhanBPMB17}
Quoc{-}Sang Phan, Lucas Bang, Corina~S. Pasareanu, Pasquale Malacaria, and
  Tevfik Bultan.
\newblock Synthesis of adaptive side-channel attacks.
\newblock In {\em IEEE Computer Security Foundations Symposium}, pages
  328--342, 2017.

\bibitem{Quisquater2001}
Jean-Jacques Quisquater and David Samyde.
\newblock {\em ElectroMagnetic Analysis (EMA): Measures and Counter-measures
  for Smart Cards}, pages 200--210.
\newblock 2001.

\bibitem{SousaD16}
Marcelo Sousa and Isil Dillig.
\newblock Cartesian hoare logic for verifying k-safety properties.
\newblock In {\em ACM SIGPLAN Conference on Programming Language Design and
  Implementation}, pages 57--69, 2016.

\bibitem{StefanBYLTRM13}
Deian Stefan, Pablo Buiras, Edward~Z. Yang, Amit Levy, David Terei, Alejandro
  Russo, and David Mazi{\`{e}}res.
\newblock Eliminating cache-based timing attacks with instruction-based
  scheduling.
\newblock In {\em European Symposium on Research in Computer Security}, pages
  718--735, 2013.

\bibitem{SungPW18}
Chungha Sung, Brandon Paulsen, and Chao Wang.
\newblock {CANAL}: A cache timing analysis framework via llvm transformation.
\newblock In {\em IEEE/ACM International Conference On Automated Software
  Engineering}, 2018.

\bibitem{TouzeauMMR17}
Valentin Touzeau, Claire Ma{\"{\i}}za, David Monniaux, and Jan Reineke.
\newblock Ascertaining uncertainty for efficient exact cache analysis.
\newblock In {\em International Conference on Computer Aided Verification},
  pages 22--40, 2017.

\bibitem{WangS17}
Chao Wang and Patrick Schaumont.
\newblock Security by compilation: an automated approach to comprehensive
  side-channel resistance.
\newblock {\em ACM {SIGLOG} News}, 4(2):76--89, 2017.

\bibitem{WangYKG08}
Chao Wang, Zijiang Yang, Vineet Kahlon, and Aarti Gupta.
\newblock Peephole partial order reduction.
\newblock In {\em International Conference on Tools and Algorithms for
  Construction and Analysis of Systems}, pages 382--396, 2008.

\bibitem{WangWLZW17}
Shuai Wang, Pei Wang, Xiao Liu, Danfeng Zhang, and Dinghao Wu.
\newblock Cache{D}: Identifying cache-based timing channels in production
  software.
\newblock In {\em USENIX Security Symposium}, pages 235--252, 2017.

\bibitem{WuGSW18}
Meng Wu, Shengjian Guo, Patrick Schaumont, and Chao Wang.
\newblock Eliminating timing side-channel leaks using program repair.
\newblock In {\em International Symposium on Software Testing and Analysis},
  2018.

\bibitem{YiYGWLZ18}
Qiuping Yi, Zijiang Yang, Shengjian Guo, Chao Wang, Jian Liu, and Chen Zhao.
\newblock Eliminating path redundancy via postconditioned symbolic execution.
\newblock {\em {IEEE} Trans. Software Eng.}, 44(1):25--43, 2018.

\bibitem{YuZW17}
Tingting Yu, Tarannum~S. Zaman, and Chao Wang.
\newblock {DESCRY:} reproducing system-level concurrency failures.
\newblock In {\em ACM SIGSOFT Symposium on Foundations of Software
  Engineering}, pages 694--704, 2017.

\bibitem{ZhangGSW18}
Jun Zhang, Pengfei Gao, Fu~Song, and Chao Wang.
\newblock {SCInfer}: Refinement-based verification of software countermeasures
  against side-channel attacks.
\newblock In {\em International Conference on Computer Aided Verification},
  2018.

\bibitem{ZhangKW15}
Naling Zhang, Markus Kusano, and Chao Wang.
\newblock Dynamic partial order reduction for relaxed memory models.
\newblock In {\em ACM SIGPLAN Conference on Programming Language Design and
  Implementation}, pages 250--259, 2015.

\end{thebibliography}

\end{document}